\documentclass[10pt, journal, compsoc]{IEEEtran}
\IEEEoverridecommandlockouts

\PassOptionsToPackage{dvipsnames}{xcolor}
\PassOptionsToPackage{switch, pagewise}{lineno}

\usepackage{amsmath,bbm}
\usepackage{graphicx}
\usepackage{textcomp}

\usepackage[frozencache]{minted}
\usepackage{tabularx}
\usepackage{amssymb}
\usepackage{graphicx}
\usepackage{textcomp}
\usepackage{fancyhdr}
\usepackage{enumitem}
\usepackage{colortbl}
\usepackage{bigstrut}
\setlist{nosep} 
\usepackage{hyperref}
\usepackage{rotating}
\usepackage{booktabs}
\usepackage{algorithm}
\usepackage[noend]{algpseudocode}
\usepackage{multicol}
\usepackage{multirow}
\usepackage{soul}
\usepackage{blkarray}
\usepackage{wrapfig}
\usepackage{caption}
\usepackage{subcaption}
\usepackage{tikz}
\usepackage{cleveref}
\usepackage{xspace}
\usepackage[normalem]{ulem}
\usepackage{tikz}
\usetikzlibrary{tikzmark}
\usepackage[framemethod=tikz]{mdframed}
\usetikzlibrary{shadows}
\usepackage{graphics}
\usepackage{adjustbox}
\usepackage{array}
\usepackage{tabularray}
\usepackage{makecell}
\usepackage{boldline}
\usepackage{listings, listings-rust, listings-algo}
\usepackage{tablefootnote}
\usepackage{proof}
\usepackage{float}
\usepackage{bussproofs}
\usepackage{mathpartir}
\usepackage{titlesec}
\usepackage{orcidlink}
\usepackage{etoolbox}

\newcommand\red[1]{\textcolor[rgb]{1.00,0.00,0.00}{#1}}

\newcommand\javadocblue[1]{\textcolor[rgb]{0.25,0.35,0.75}{{#1}}}

\newcommand\purple[1]{\textcolor[rgb]{0.22,0.0,0.44}{#1}}

\definecolor{blueish}{RGB}{250, 250, 255}
\definecolor{greenish}{RGB}{250, 255, 250}
\definecolor{redish}{RGB}{255, 200, 200}

\definecolor{highlight}{RGB}{175, 255, 100}
\definecolor{gray01}{gray}{.98}
\definecolor{gray05}{gray}{0.95}
\definecolor{gray08}{gray}{0.92}
\definecolor{gray10}{gray}{0.90}
\definecolor{gray12}{gray}{0.88}
\definecolor{gray15}{gray}{0.85}
\definecolor{gray18}{gray}{0.82}
\definecolor{gray20}{gray}{0.80}
\definecolor{gray25}{gray}{0.75}
\definecolor{gray30}{gray}{0.70}
\definecolor{gray35}{gray}{0.65}
\definecolor{gray40}{gray}{0.60}
\definecolor{gray45}{gray}{0.55}
\definecolor{gray50}{gray}{0.50}
\definecolor{gray55}{gray}{0.45}
\definecolor{gray60}{gray}{0.40}
\definecolor{gray65}{gray}{0.35}
\definecolor{gray70}{gray}{0.30}
\definecolor{gray75}{gray}{0.25}
\definecolor{gray80}{gray}{0.20}
\definecolor{gray85}{gray}{0.15}
\definecolor{gray90}{gray}{0.10}
\definecolor{gray95}{gray}{0.05}
\def\thesubsubsectiondis{\unskip\arabic{subsubsection})}
\newcommand{\bi}{\begin{itemize}[leftmargin=*, wide=0pt]}
\newcommand{\ei}{\end{itemize}}
\newcommand{\beq}{\begin{equation}}
\newcommand{\eeq}{\end{equation}}
\newcommand{\be}{\begin{enumerate}[leftmargin=*, wide=0pt]}
\newcommand{\ee}{\end{enumerate}}

\newcommand{\sysname}{\textsc{Yuga}\xspace}
\newcommand{\tool}{\textsc{Yuga}\xspace}

\newmdenv[
    tikzsetting= {fill=gray!10},
    skipabove=0.33em,
    skipbelow=0.33em,
    linewidth=0.5pt,
    innerleftmargin=4pt,
    innerrightmargin=4pt,
    innertopmargin=2pt,
    innerbottommargin=2pt,
    linecolor=gray95,
    roundcorner=2pt, 
    shadow=true,
    shadowsize=2pt,
    shadowcolor=black
]{myshadowbox}

\newcommand{\etal}{\hbox{\emph{et al.}}\xspace}
\newcommand{\eg}{\hbox{\emph{e.g.,}}\xspace}
\newcommand{\ie}{\hbox{\emph{i.e.,}}\xspace}

\newcommand{\la}{\textsc{LA}\xspace}

\newcommand{\las}{\textsc{LA}s\xspace}
\newcommand{\Las}{\textsc{LA}s\xspace}
\newcommand{\laparan}{(\textsc{LA}s)\xspace}

\newcommand{\rust}{Rust\xspace}

\definecolor{javared}{rgb}{0.6,0,0} 
\definecolor{javagreen}{rgb}{0.25,0.5,0.35} 
\definecolor{javapurple}{rgb}{0.5,0,0.35} 
\definecolor{javadocblue}{rgb}{0.25,0.35,0.75} 

\newcommand{\TT}[1]{{\javadocblue{\texttt{\small#1}}}}

\newcommand{\rebuttal}[2]{#2}

\newcommand{\minorrevision}[2]{#2}

\crefformat{footnote}{#2\footnotemark[#1]#3}

\algdef{SE}[DOWHILE]{Do}{doWhile}{\algorithmicdo}[1]{\algorithmicwhile\ #1}%

\def\BibTeX{{\rm B\kern-.05em{\sc i\kern-.025em b}\kern-.08em
    T\kern-.1667em\lower.7ex\hbox{E}\kern-.125emX}}
\begin{document}

\title{\sysname{}: Automatically Detecting Lifetime Annotation Bugs in the \rust Language}

\author{Vikram Nitin\orcidlink{0009-0004-8620-8255}, Anne Mulhern\orcidlink{0009-0005-8895-0688}, Sanjay Arora\orcidlink{0009-0003-4731-9959}, and Baishakhi Ray\orcidlink{0000-0003-3406-5235} 
\IEEEcompsocitemizethanks{\IEEEcompsocthanksitem Vikram Nitin and Baishakhi Ray are with the Department of Computer Science, Columbia University, New York, USA. E-mail: vikram.nitin@columbia.edu, rayb@cs.columbia.edu
\IEEEcompsocthanksitem Anne Mulhern and Sanjay Arora are with RedHat Research, USA. E-mail: \{amulhern, saarora\}@cs.columbia.edu}
}


\markboth{Journal of \LaTeX\ Class Files,~Vol.~14, No.~8, August~2021}%
{Shell \MakeLowercase{\textit{et al.}}: A Sample Article Using IEEEtran.cls for IEEE Journals}

\maketitle

\thispagestyle{plain}
\pagestyle{plain}

\begin{abstract}
The Rust programming language is becoming increasingly popular among systems programmers due to its efficient performance and robust memory safety guarantees. 
Rust employs an ownership model to ensure these guarantees by allowing each value to be owned by only one identifier at a time. It uses the concept of borrowing and lifetimes to enable other variables to temporarily borrow values.

Despite its benefits, security vulnerabilities have been reported in Rust projects, often attributed to the use of "unsafe" Rust code. These vulnerabilities, in part, arise from incorrect lifetime annotations on function signatures. However, existing tools fail to detect these bugs, primarily because such bugs are rare, challenging to detect through dynamic analysis, and require explicit memory models.

To overcome these limitations, we characterize incorrect lifetime annotations as a source of memory safety bugs and leverage this understanding to devise a novel static analysis tool, \tool, to detect potential lifetime annotation bugs. \tool uses a multi-phase analysis approach, starting with a quick pattern-matching algorithm to identify potential buggy components and then conducting a flow and field-sensitive alias analysis to confirm the bugs. We also curate new datasets of lifetime annotation bugs. \tool successfully detects bugs with good precision on these datasets, and we make the code and datasets publicly available.

\end{abstract}


\section{Introduction}

The \rust programming language is gaining increasing popularity among systems programmers for its efficient performance while providing strong memory safety, thread safety, and type checking. It uses a C-like syntax that allows low-level access, enabling it to be used on embedded devices, to write kernel code, to build performance-critical systems, and so on. At the same time, it ameliorates many of the pitfalls of C, like pointer errors and concurrency errors. For these reasons, Google recently extended support for developing the Android OS in \rust \cite{google-2021a}, and \rust is officially supported in the Linux kernel as of Linux 6.1 \cite{vaughan-nichols-2022}.



\smallskip
\noindent
\textbf{\rust memory safety model.} While working within the secure confines of \rust, we enjoy robust safety guarantees.
Central to these assurances is \rust's \textit{Ownership} model\textemdash a set of rules dictating memory management within a \rust program. 
Ownership guarantees that each value and its associated memory are owned by exactly one identifier at any point in the program. Modern programming languages handle memory usage during execution in different ways---some rely on garbage collection (e.g., Java) and others on manual allocation and deallocation (e.g., C). \rust takes an alternative approach with the concept of ownership, reinforced by compiler-verified rules.
Yet, none of these rules significantly slow down the run-time of \rust programs.

\begin{figure*}[t]
    \centering
    \begin{subfigure}{0.48\textwidth}
    \includegraphics[width=\textwidth]{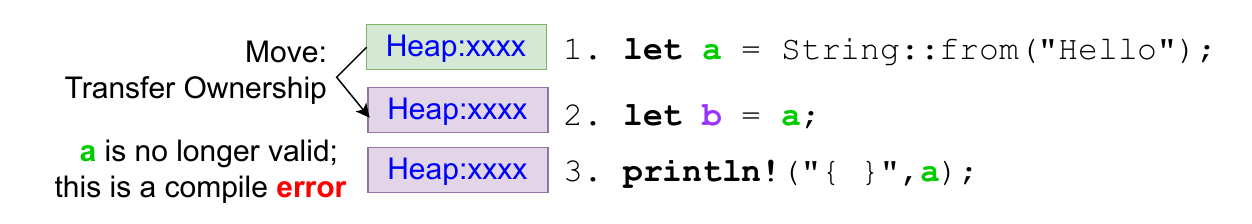}
    \caption{\footnotesize{\textbf{Ownership.} The read of \TT{\color{green}{a}} at Line 3 causes the program to not compile because  \TT{\purple{b}} has taken ownership of the \TT{String} value and so \TT{\color{green}{a}} is not live.}}
    \label{fig:ownership}   
    \end{subfigure}
   ~
    \begin{subfigure}{0.48\textwidth}
    \includegraphics[width=\textwidth]{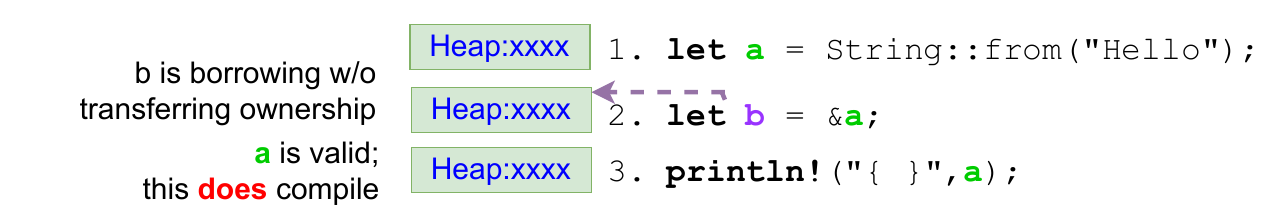}
    \caption{\footnotesize{\textbf{Borrowing}. When the \TT{String} is borrowed by \TT{\purple{b}} (Line 2), \TT{\color{green}{a}} still retains ownership of the \TT{String} and so of the heap region \TT{XXXX} which the \TT{String} owns.}}
    \label{fig:mutability}
    \end{subfigure} 
    \caption{\small{\textbf{Ownership and Borrowing.} The color of an identifier indicates the color of the heap region it owns.}}
    \label{fig:scope}
\end{figure*}

However, the strict ownership policy---each value owned by only one identifier at any given time, is often overly limiting. Thus, \rust allows \textit{Borrowing}---another identifier may temporarily borrow a value under certain conditions by creating a reference to that value. Consider the statement \TT{a = \&b}, here \TT{a} is borrowing the value \TT{b}.
\rust further introduces the concept of \textit{Lifetimes} to restrict a borrow's scope---a lifetime is a region of the source code in which the borrowing identifier is live; the borrowing identifier must not outlive the borrowed value (\eg~  \TT{a} cannot outlive \TT{b} in the above example). 

The \rust compiler, specifically its \textit{Borrow Checker}, ensures the validity of all borrows by tracking variable lifetimes. Often the checker can infer the lifetime of a variable without assistance.
However, for certain cases related to inter-procedural reasoning, 
the compiler requires explicit \textit{lifetime annotations} on function signatures to guide the Borrow Checker. For example, the type \TT{\&'a i32} denotes a borrow to an integer value, where the lifetime of the borrow is 
parameterized by the lifetime annotation \TT{'a}.
These annotations on the function signature provide valuable information to the compiler, enabling it to verify certain memory safety properties without even looking at the function body.

\smallskip
\noindent
\textbf{Violation of memory safety guarantees.}
Despite the strong safety guarantees promised by \rust, an increasing number of security vulnerabilities have been reported in \rust projects. The number of vulnerabilities reported to the RUSTSEC Database \cite{rustsec} during 2018 was only 22; in 2022, there were 91. The majority of these vulnerabilities arise from the use of a language feature called \TT{unsafe} \rust \cite{xu2020memory}---\rust's memory safety guarantees can be too restrictive for some applications, so \rust provides an ``escape hatch'' in the form of \TT{unsafe} \rust, which allows writing code that bypasses some 
memory safety checks.
The added flexibility comes at the cost of potential memory safety-related bugs.

Thus, while using \TT{unsafe} \rust, it is the developer's responsibility to ensure the code is correct---a single mistake may jeopardize the whole program's safety guarantees and introduce vulnerability. One of the causes of such vulnerabilities is incorrect lifetime annotations \laparan on function signatures.
Deciding the correct \las for a variable is non-trivial---it requires expert knowledge of how the value will propagate to subsequent borrows and pointers, often through deeply nested structures and function calls. Further, lifetimes in \rust are a non-intuitive concept even for experienced programmers, as there is no analogous concept in any other popular programming language (to the best of our knowledge). A one-token mistake in specifying correct \las can lead to security vulnerabilities\cite{futures_rs}, as we see from a manual inspection of reported \rust security vulnerabilities.

\textbf{Limitation of Existing Approaches.} Checking the safety of unsafe \rust is an active area of research \cite{baranowski2018verifying, matsushita2021rusthorn, matsushita2022rusthornbelt, toman2015crust, almohri2018fidelius, rivera2021keeping}, with multiple reported vulnerabilities \cite{rustsec, xu2020memory}, and automated tools that perform static \cite{bae2021rudra, li2021mirchecker, cui2021safedrop, rustclippy}, and dynamic \cite{miri, cargofuzz} analysis to check for vulnerabilities.
However, none of these existing tools can detect memory-safety bugs due to incorrect \las. 
First of all, \la bugs are rare and typically only surface during execution in specific edge cases. Thus, detecting them through dynamic analysis is non-trivial. Additionally, they tend to be deeply embedded within a program, and analyzing them requires accurate modeling of program memory. Thus, it becomes challenging to study using standard static analyzers or symbolic execution tools. 
Even, more mature hybrid vulnerability detection tools designed for other languages like C/C++  cannot be applied here as \la is a new concept introduced in \rust.

To this end, we propose a novel static analysis tool to check correct memory usage in unsafe \rust code. 
In particular, we focus on detecting incorrect lifetime annotations \laparan as they often lead to security vulnerabilities.

\textbf{Our approach:} 
We have developed and implemented a novel static analysis tool called \tool, capable of detecting incorrect \la bugs. 
We employ a multi-phase analysis approach that starts with a pattern-matching algorithm to quickly identify the potential buggy components. 
Subsequently, we perform a flow and field-sensitive alias analysis, focusing only on the potential buggy components to confirm the bugs. 
By adopting this multi-phase analysis, we effectively address the well-known scalability challenges associated with alias analysis, while still achieving reasonably high precision.


In particular, \tool operates in four stages. First, it propagates the \las from a function parameter to internal references, pointers, and fields, using an AST-like representation. In the second stage, it identifies any potential mismatches in \las between the function arguments and return values. These first two steps are challenging since they involve analyzing relationships between program variables through nested pointers and structures. When a mismatch is found, \tool flags the corresponding pairs of values as potentially buggy. 
In the third stage, \tool conducts a flow and field-sensitive alias analysis to confirm the interactions between the source and target, further validating the presence of bugs. Up to this point, the analysis is intra-procedural. In the final stage, \tool employs specific shallow filters based on common inter-procedural knowledge to reduce false positives, improving the overall precision.

\textbf{Results.} We evaluate \tool using a dataset of known security vulnerabilities, and a synthesized vulnerability dataset. We find that \tool can detect \la bugs with \rebuttal{}{87.5\%} precision \rebuttal{}{on this dataset of known vulnerabilities}. None of the existing static analysis tools we evaluated could detect \textit{any} of these vulnerabilities.
We also ran \tool on 372 \rust libraries, and \tool generated \rebuttal{}{85} reports. Of these, 3 were confirmed to be exploitable vulnerabilities, and the rest were ``code smells'' that are not exploitable but indicate potential for a vulnerability to arise in future development.



\textbf{Contributions:} Our contributions in this paper are: 
\be
\item We are the first to characterize incorrect lifetime annotations that can lead to memory safety bugs. 
\item We propose \sysname{}, a novel static analysis technique, that can automatically scan \rust projects at scale and detect potential \la bugs that can lead to vulnerabilities. 
\item We evaluate \sysname{} in a variety of settings and show that it can successfully detect \la bugs with good precision. 
To the best of our knowledge, there are no existing baseline static analyzers that can find these bugs. 

\item We contribute two new datasets of lifetime annotation bugs, one created from \rust security vulnerabilities, and the other synthetic.  We make the code and dataset publicly available at \url{https://github.com/vnrst/Yuga}, \rebuttal{[Reviewer 3: Dataset GitHub link]}{and \url{https://github.com/vnrst/rust-lifetime-bugs} respectively.}

\ee


The rest of the paper is structured as follows. In~\Cref{sec:background}, we discuss the features of \rust and lifetime annotations. Then, we define three patterns of lifetime annotation bugs in~\Cref{sec:patterns}.~\Cref{sec:methodology} presents the design of \sysname{}, our tool to detect lifetime annotations bugs. We evaluate the performance of \sysname{} in~\Cref{sec:setup,sec:rqs}, present related work in~\Cref{sec:related_work}, and conclude in~\Cref{sec:conclusion}.
\section{Background}
\label{sec:background}

\begin{figure*}[!th]
\lstset{escapeinside={<@}{@>}}
 \centering
  \begin{subfigure}{0.35\textwidth}
    \includegraphics[width=\textwidth]{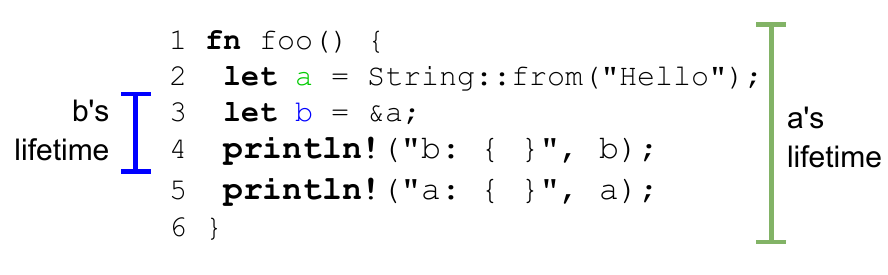}
    \caption{\small{The lifetime of the borrow, \ie ~\TT{b}'s lifetime, must not extend beyond the lifetime of the borrowed value \TT{a}.}}
    \label{fig:lifetime-a}
    \end{subfigure} 
    ~~\hspace{5pt}
    \begin{subfigure}{0.25\textwidth}
    \lstset{xleftmargin=.11\columnwidth}
    \begin{lstlisting}[language=Rust, frame=none, style=colouredRust, basicstyle=\footnotesize\ttfamily, numbers=left]
let b : &i32;
{   let a : i32 = 5;
    b = &a;
} println!("{}", b);
\end{lstlisting}
    \caption{\small{The compiler will reject this program since \TT{b}'s lifetime extends till Line 4, but \TT{a}'s lifetime is terminated by the closing brace on Line 3.}}
    \label{fig:lifetime-b}
    \end{subfigure} 
    ~~\hspace{5pt}
    \begin{subfigure}{0.3\textwidth}
    \lstset{xleftmargin=.09\columnwidth}
    \begin{lstlisting}[language=Rust, frame=none, style=colouredRust, basicstyle=\footnotesize\ttfamily, numbers=left]
let mut a = 5;
let ref1 = &a;
let ref2 = &mut a;
println!("{}", ref1);
\end{lstlisting}
    \caption{\small{The compiler will reject this program because the lifetimes of immutable borrow \TT{ref1} (Lines 2-4) and mutable borrow \TT{ref2} (Line 3) are not disjoint.}}
    \label{fig:lifetime-c}
    \end{subfigure}
\caption{\small{\textbf{Lifetimes in \rust}.}}
\label{fig:lifetime}
\end{figure*}

This section provides an overview of \rust's ownership model and the memory safety properties associated with it. 
~\Cref{sec:safe_rust} explains how \rust ensures memory safety via its ownership model. ~\Cref{sec:unsafe_rust} discusses how \rust can be used in an \textit{unsafe} manner.


\subsection{Rust Memory Safety Model}
\label{sec:safe_rust}

\subsubsection{Ownership.}
\rust's ownership model ensures that a memory location has exactly one owner at any point in the program, and the associated memory is freed when the owner goes out of scope. For example, in ~\Cref{fig:ownership}, at Line 1, the identifier \TT{a} is assigned a \TT{String} value. The characters of the \TT{String} are stored on the heap, and \TT{a} takes ownership of this heap-allocated memory. At Line 2, \TT{a} is assigned to \TT{b}. This \textit{move} transfers the ownership of the heap-allocated memory to \TT{b}. Once \TT{b} takes ownership of this memory, Rust no longer considers a use of \TT{a} to be valid, as a memory location must have exactly one owner. So the print statement on Line 3 will result in a compile error.

\subsubsection{Borrowing.}
A value can be \textit{borrowed} by creating a reference to that value using the `\TT{\&}' symbol. A reference can be mutable or immutable, depending on whether one can modify the borrowed value through the reference or not, respectively. Mutable references are denoted as \TT{\&mut}.
Figure \ref{fig:mutability} illustrates borrowing. At Line 1, \TT{String} \TT{"Hello"} is assigned to an identifier \TT{a}, and \TT{a} takes ownership of the \TT{String}. At Line 2, \TT{b} merely borrows the \TT{String} (i.e., it does not take ownership) by taking a reference to \TT{a}. \TT{a} retains ownership of the heap-allocated memory associated with the \TT{String}. Consequently, the compiler does not reject the use of \TT{a} at Line 3, and the program compiles.

\begin{figure*}[ht]
\centering
\begin{subfigure}{0.43\textwidth}
\lstset{xleftmargin=0\textwidth, xrightmargin=0.3\textwidth, escapeinside={<@}{@>}}
\begin{lstlisting}[language=Rust, frame=none, style=colouredRust, basicstyle=\footnotesize \ttfamily]
struct Foo<'a> {
    x: *const String,
    ..
}
fn bar<'a,'b>(arg1: &<@\red{'a}@> String, arg2: &'b String)
        -> Foo<<@\red{'a}@>> {
    Foo{x: arg2 as *const String, ..}
}
\end{lstlisting}
\caption{The function returns a pointer that derives from \TT{arg2}, but the annotation indicates that it uses \TT{arg1}.}
\label{fig:encapsulating}
\end{subfigure}
~~\hspace{10pt}
\begin{subfigure}{0.48\textwidth}
\includegraphics[width=0.9\columnwidth]{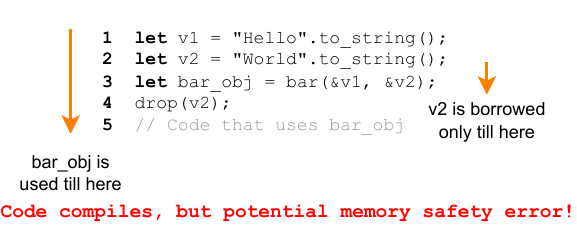}
    \caption{The string \TT{v2} is dropped on Line 4, and \TT{bar\_obj} will contain a pointer to freed memory. This can cause a use-after-free error.}
    \label{fig:invalid}
\end{subfigure}
\caption{Incorrect lifetime annotations on functions can cause memory safety errors.} 
\end{figure*}

\subsubsection{Lifetimes.}
\label{sec:back_lifetime}
A \rust lifetime is similar to a scope\textemdash it is a statically determined region of a program. However,
it is not always defined by a lexical marker such as a closing curly brace \cite{nonlexicallifetimes}. For example, in~\Cref{fig:lifetime-a}, \TT{b}'s lifetime is just Lines 3-4.

If a reference is assigned to an identifier, \eg~\TT{b = \&a}, then the lifetime of the identifier is called the \textit{lifetime of the borrow}. Two rules constrain the lifetimes of borrows:
\be
    \item[\textbf{Rule 1:}] \textit{The lifetime of a borrow must not extend beyond the lifetime of the borrowed value. In other words, the borrowed value must \text{\textnormal{``outlive''}} the borrow.}
    \item[\textbf{Rule 2:}] 
    \textit{If there are multiple borrows to the same value, every mutable borrow must have a lifetime which is disjoint from that of every other borrow, whether mutable or immutable.}
\ee

\rust's Borrow Checker ensures that these rules are not violated. The Borrow Checker must assign a lifetime to each identifier such that the assignment satisfies these rules. Rule 1 helps prevent 
some common memory errors as use-after-free and double-free. 
Rule 2 promotes thread safety by preventing concurrent writes and deadlocks. 
It further allows the compiler to perform numerous low-level optimizations, e.g., eliminating unnecessary reads from memory.

~\Cref{fig:lifetime} illustrates the concept of lifetimes. In~\Cref{fig:lifetime-a},
\TT{a} owns the \TT{String} while \TT{b} borrows it. 
By Rule 1, it is essential that \TT{a} outlives \TT{b}, and this is satisfied.
In contrast, in ~\Cref{fig:lifetime-b}, \TT{b}'s lifetime extends until line 4, but \TT{a}'s lifetime extends only until line 3. So the Borrow Checker rejects this program.
In~\Cref{fig:lifetime-c} the lifetime of \TT{ref2} (Line 3), which borrows \TT{a} mutably, is not disjoint from the lifetime of \TT{ref1} (Line 2-4) which also borrows \TT{a}. So the Borrow Checker will reject this program by Rule 2.


\noindent
\textit{Lifetime Annotations on Functions.}
So far, we have discussed how the compiler infers lifetimes when a reference is assigned directly to an identifier, like \TT{b = \&a}. Now consider a function call where a parameter or return type is a reference.
In the simple case of a function with one reference parameter and one returned reference, e.g., \lstinline[language=Rust, frame=none, style=colouredRust, basicstyle=\small\ttfamily]!fn foo(x: &String) -> &String!, the lifetimes of the argument and the returned reference are calculated as if the invocation of the function is just a use of the argument. So,
\begin{lstlisting}[language=Rust, frame=none, style=colouredRust, basicstyle=\footnotesize\ttfamily]
  let b = String::from("Hello");
  let a = foo(&b);
\end{lstlisting}
is equivalent to
\begin{lstlisting}[language=Rust, frame=none, style=colouredRust, basicstyle=\footnotesize\ttfamily]
  let b = String::from("Hello");
  let a = &b;
\end{lstlisting}
for the purpose of inferring lifetimes. In both cases, the lifetime of \TT{a} is considered as the lifetime of the borrow, and \TT{b} must outlive \TT{a}. Consider, however, a function that takes more than one parameter.
\begin{lstlisting}[language=Rust, frame=none, style=colouredRust, basicstyle=\footnotesize\ttfamily]
  fn foo(x: &String, y: &String) -> &String
\end{lstlisting}
The returned reference must not outlive the value that it's borrowing, by Rule 1. But it's not clear what that borrowed value is. It could be derived from any one of the two input arguments. In this situation, the compiler requires the developer to add \las.
\begin{lstlisting}[language=Rust, frame=none, style=colouredRust, basicstyle=\footnotesize\ttfamily]
  fn foo<'a,'b>(x: &'a String, y: &'b String)
            -> &'b String { y }
\end{lstlisting}
Here, \TT{'a} and \TT{'b} are lifetime annotations \laparan that represent the actual lifetimes of the two input borrows. In this example, the developer has chosen to declare that the
returned reference derives from the second argument, by placing the lifetime parameter \TT{'b} on
second parameter and on the return type. The compiler will reject the function if the returned reference is not a reference to some part of \TT{y}.

\Las are only a syntactic representation of a concrete lifetime, \ie~a region of code. However for conciseness, we shall sometimes use expressions like ``the lifetime \TT{'a}'' to mean ``the lifetime associated with the \la~\TT{'a}''.

\noindent
\textit{Lifetime Annotations on Structures.} 
\Las can be used only in two specific contexts - the first is in function signatures, as shown above, and the second is in the declaration of algebraic data types like structures. If a structure contains a reference as one of its fields, then that reference \textit{must} have an associated \la. Further, that same annotation \textit{must} appear as part of the structure type. For example,
\lstinline[language=Rust, frame=none, style=colouredRust, basicstyle=\footnotesize\ttfamily]
!struct Foo<'a>{x: &'a i32}!
is a structure \TT{Foo} containing a reference with \la \TT{'a}, so \TT{<'a>} is part of its type.
The significance of this annotation is that \TT{'a} is the lifetime of the borrow, \textit{so the structure object must not outlive \TT{'a}}.

\subsection{Unsafe \rust}
\label{sec:unsafe_rust}

The framework discussed above guarantees memory safety~\cite{bae2021rudra,nomicon}. However, these guarantees are inherently conservative and prevent access to many useful constructs often required for systems programming, like low-level memory manipulation, external function calls, etc. 
To support such functionalities, \rust has an extended language called unsafe \rust, which allows some additional features that bypass some of the safety guarantees of safe \rust. 
In this connection, we discuss the \TT{unsafe} keyword, \textit{raw pointers}, and \textit{lifetime annotation}. 
Evans \etal~\cite{evans2020rust} provide a more comprehensive discussion of unsafe \rust.

\subsubsection{The \TT{unsafe} Keyword.}
The \TT{unsafe} keyword is used to distinguish regions containing unsafe \rust from the rest of a program. If a function is marked unsafe then the body of the function may use unsafe Rust and a function caller must treat the function as unsafe. If a block within a function is marked unsafe, then the developer must
ensure the the function
provides the same guarantees of safe \rust when it is used.
Just one function that violates this contract may introduce errors that break the memory safety guarantees of the entire program.

\subsubsection{Raw Pointers.}
Raw pointers are similar to pointers in C - they obey none of the guarantees that references in \rust are expected to obey. For example, one can have mutable and immutable raw pointers to the same location. The memory location it points to  need not live as long as a raw pointer itself, which means that a raw pointer is not guaranteed to point to valid memory. A raw pointer type does not allow \las. 
Although \textit{creating} and manipulating a raw pointer is permitted in safe Rust, any \textit{dereference} of a raw pointer must be done in an \TT{unsafe} block or function.

\subsubsection{Unsafe Code and Lifetime Annotations.}
\label{sec:back_life_anno}

Although raw pointers themselves don't have \las, they can be contained within \textit{structure objects}, and these structure objects can have associated \las. In ~\Cref{fig:encapsulating}, function \TT{bar} returns a structure containing a raw pointer. In \TT{bar}'s signature the placement of the \la \textcolor{red}{\TT{'a}} in \TT{x: \&\textcolor{red}{'a}} and \TT{Foo<\textcolor{red}{'a}>} informs the compiler that the
structure object is tied to the lifetime of \TT{x}.


Although the code in the example contains no \TT{unsafe} regions, it still contains a bug which the compiler did not detect; the \la on the result type should be \TT{b}, not \TT{a}, because it is \TT{y}, not \TT{x}, which is assigned to the \TT{inner} field of the returned structure. This can pose a problem as shown in Fig \ref{fig:invalid}. The borrow lifetime \TT{'a} is tied to \TT{bar\_obj}, and extends till Line 5. But the borrow lifetime \TT{'b} extends only till line 3, and so \textit{\TT{v2} is no longer borrowed after line 3}. It can be dropped on Line 4 without a compile error. Now the \TT{inner} field of \TT{bar\_obj} points to freed memory, and will result in a use-after-free error if dereferenced in an \TT{unsafe} block.

\rebuttal{[Reviewer 2: Role of unsafe superpowers]}{At a high level, our analysis deals with raw pointers and references, and tries to identify situations where the values that they point to get freed. Raw pointers go hand-in-hand with unsafe code, because most operations involving raw pointers, like dereferencing and freeing, require \TT{unsafe}.
}

\section{Methodology}
First we describe our process of classifying and characterizing \la bugs in~\Cref{sec:patterns}. Then we use these patterns to develop \sysname{}, as described in~\Cref{sec:methodology}.

\subsection{Characterizing Lifetime Annotation Bugs}
\label{sec:patterns}

\noindent
Within a function, 
some values may be transferred between two arguments (\textbf{arg-arg}) or between an argument and returned value (\textbf{arg-return}). 
When the transfer of a value happens, the \las on the \textit{target} of the transfer should be consistent with the \textit{source} of the transferred value. Otherwise, it leads to a \la bug. 

Arguments and return types are the only places where \las can occur in a function signature. Therefore, all \la bugs fall into one of two categories: (i) \textit{mismatched arg-arg} or (ii) \textit{mismatched arg-return}. This claim is validated by an initial study of \la bugs reported to the RustSec Security Advisory Database (see~\Cref{sec:rqs} for details).
Such incorrect \las may lead to memory safety bugs, like:
\be
\item \textbf{Use-after-free (UAF):} Such bugs occur when we attempt to access memory that has been freed or de-allocated. 
\item \textbf{Non-exclusive mutability (NEM):} These occur when two mutable borrows co-occur within the same lifetime. 
\ee

\noindent
We describe these bugs in the rest of the section in detail. 

\subsubsection{Use-after-free Bugs}
\label{sec:use-after-free}

Let's say a function takes a value that is associated with a borrow lifetime \TT{'a} and assigns it to a location where it is associated with a borrow lifetime \TT{'b}. If \TT{'b} is longer than \TT{'a}, a UAF error could occur (violation of Rule 1 in~\Cref{sec:back_lifetime}). This can be caused by arg-arg or arg-return assignment.

\noindent
\textbf{Arg-Return Assignment:}
\begin{table}[!th]
  \caption{\textbf{\small{Lifetime Annotation Bugs of Different Types. The incompatible lifetime annotations are highlighted in red.}}}
    \label{tab:ex_bugs}
    \centering
    \begin{tabular}{p{0.9\columnwidth}}
    \toprule
     \textbf{Example 1.} RUSTSEC-2021-0130: \textbf{UAF} bug in the \TT{lru} crate caused by assigning a value from an argument to return (\textbf{arg-return}) \\
    \midrule
    \lstset{xleftmargin=0.05\columnwidth, xrightmargin=0\columnwidth, escapeinside={<@}{@>}}
\setlength{\fboxsep}{2.2pt}
\sethlcolor{gray20}
\begin{lstlisting}[language=Rust, frame=none, style=colouredRust, basicstyle=\scriptsize \ttfamily, numbers=left]
pub struct LruEntry<K, V> {
    next: *mut LruEntry<K, V>, ..
}
pub struct LruCache<K, V> {
    head: *mut LruEntry<K, V>, ..
}
pub struct Iter<'a, K: 'a, V: 'a> {
    ptr: *const LruEntry<K, V>, ..
}
impl<K, V, S> LruCache<K, V, S> {
    ...
<@\colorbox{gray15}{\  \textbf{pub fn} iter<\textcolor{red}{'a}>(\&\textcolor{red}{'\_} \textbf{self}) -> Iter<\textcolor{red}{'a}, K, V> \{ }@>
        Iter {
            ptr: unsafe { (*self.head).next },
            ..
        }
    }
}
\end{lstlisting} \\ 
\bottomrule
\\
\textbf{Example 2.} RUSTSEC-2021-0128: \textbf{UAF} bug in the \TT{rusqlite} crate caused by assigning a value from one argument to another argument (\textbf{arg-arg}).\\
\midrule
\lstset{xleftmargin=0.05\columnwidth, xrightmargin=0\columnwidth, escapeinside={<@}{@>}}
\sethlcolor{gray20}
\setlength{\fboxsep}{2.2pt}
\begin{lstlisting}[language=Rust, frame=none, style=colouredRust, basicstyle=\scriptsize \ttfamily, numbers=left]
pub struct InnerConnection {
  db: *mut ffi::sqlite3,
  ..
}
pub struct Connection {
  db: RefCell<InnerConnection>
}
impl Connection {
    ..
<@\colorbox{gray15}{\hspace{8pt}\textbf{pub fn} update\_hook<\textcolor{red}{'c},F>(\&\textcolor{red}{'c} \textbf{self},hook:Option<F>)}@>
<@\colorbox{gray15}{\hspace{16pt}where}@>
<@\colorbox{gray15}{\hspace{16pt}F: .. + \textcolor{red}{'c}}@>
  {
    self.db.borrow_mut().update_hook(hook);
  }
}
\end{lstlisting}\\
\bottomrule
\\
  \textbf{Example 3.} RUSTSEC-2020-0023: \textbf{NEM} bug in the \TT{rulinalg} crate caused by assigning a value from an argument to return (\textbf{arg-return}) \\
\midrule
\lstset{xleftmargin=0.05\columnwidth, xrightmargin=0\columnwidth, escapeinside={<@}{@>}}
\sethlcolor{gray}
\setlength{\fboxsep}{2pt}
\begin{lstlisting}[language=Rust, frame=none, style=colouredRust, basicstyle=\scriptsize \ttfamily, numbers=left]
pub struct MatrixSliceMut<'a, T: 'a> {
    ptr: *mut T, ..
}
pub struct RowMut<'a, T: 'a> {
    row: MatrixSliceMut<'a, T>
}
impl<'a, T: 'a> RowMut<<@\textcolor{red}{'a}@>, T> {
  /// Returns the row as a slice.
<@\colorbox{gray15}{\hspace{8pt}\textbf{pub fn} raw\_slice\_mut(\&\textcolor{red}{'\_} \textbf{mut self})->\&\textcolor{red}{'a} \textbf{mut} [T] \{}@>
    unsafe { from_raw_parts_mut(self.row.ptr, ..) }
  }
}
\end{lstlisting} \\
\bottomrule
    \end{tabular}
\end{table}
Example 1 in \Cref{tab:ex_bugs} shows an example of a real-world UAF bug caused by arg-return \rebuttal{}{assignment}. On Line 12, the function \TT{iter}, takes an argument \TT{self}, and by following the fields of each structure, we see that \TT{(*self.head).next} is a raw pointer to a value of type \textcolor{red}{\TT{LruEntry}}, and it is associated with the \la \textcolor{red}{\TT{'\_}}. On Line 14, it is assigned to the \TT{ptr} field of the returned \TT{Iter} object, where it is associated with the \la \textcolor{red}{\TT{'a}} on \TT{Iter<'a, ..>}.
The \las do not correctly specify the relationship between the lifetimes of the two values, so the compiler will accept a program that uses the returned \TT{Iter} object even after \TT{(*self.head).next} has been freed, leading to a UAF.


\noindent
\textbf{Arg-Arg Assignment:} There is a subtle difference in checking \las for arg-arg assignments. In Example 2 in~\Cref{tab:ex_bugs}, for every call of the \TT{update\_hook} function, \TT{hook} must live at least as long as \TT{self}. However, the \las do \textit{not} enforce this requirement and could lead to a UAF. Both \TT{hook} and \TT{self} have the same 
\la \textcolor{red}{\TT{'c}}, which means that they both live \textit{at least} as long as \TT{'c}. But this does not mean that \TT{hook} lives as long as \TT{self}.



\subsubsection{Non-Exclusive Mutability Bugs}
\label{sec:mutability}

Incorrect \las can allow the lifetimes of multiple mutable borrows to overlap (i.e., violate Rule 2 in~\Cref{sec:back_lifetime}), leading to NEM. This kind of error can be caused by arg-return or arg-arg assignment. However, in our empirical analysis (as we shall describe in~\Cref{sec:registry}), we did not observe any NEM bugs caused by arg-arg assignment. We therefore exclude them from the scope of this paper, and leave it to future work.

\noindent
Example 3 in~\Cref{tab:ex_bugs} shows a real-world example of a NEM bug caused by arg-return assignment. On Line 9, \TT{raw\_slice\_mut}, takes a mutable borrow to \TT{self}. The mutable pointer \TT{self.row.ptr} of type \TT{*mut T} is re-borrowed for the lifetime \textcolor{red}{\TT{'\_}}. This same value is then returned through a borrow with lifetime \textcolor{red}{\TT{'a}}. It is possible for \textcolor{red}{\TT{'a}} to be longer than \textcolor{red}{\TT{'\_}}, so \TT{raw\_slice\_mut} can be called multiple times to create multiple mutable borrows to the same \TT{*(self.row.ptr)} value.


\subsection{Algorithm Design}
\label{sec:methodology}

\begin{figure}[t]
\centering
\begin{subfigure}{0.98\columnwidth}
\lstset{xleftmargin=0.15\columnwidth, xrightmargin=0.05\columnwidth, escapeinside={<@}{@>}}
\begin{lstlisting}[language=Rust, frame=nones, style=colouredRust, basicstyle=\footnotesize \ttfamily]
struct Foo<'a> { x: *mut String,
                 w: &'a mut i32 }
struct Bar { y: String,
             z: *mut i32 }
             
fn bar<'a,'b>(arg1: &'a mut i32,
              arg2: &'b mut Bar)
        -> Foo<'a> {
    let ret = Foo{ x: &mut (*arg2).y,
                   w: arg1 };
    ret
}
\end{lstlisting}
\end{subfigure}
{\small
\begin{subfigure}{0.98\columnwidth}
\centering
    \renewcommand{\arraystretch}{0.88}
        \begin{tabular}{c|c|c}
        \toprule
            \footnotesize\textbf{Value} & \footnotesize\textbf{Type} & \footnotesize\textbf{Borrowed For} \\
            \midrule
            \texttt{\footnotesize arg1} & \texttt{\footnotesize \&'a \textbf{mut} i32} & \texttt{\footnotesize None} \\
            \rowcolor{gray10}
            \tikzmark{a2}
            \texttt{\footnotesize *arg1} & \texttt{\footnotesize i32} & \texttt{\footnotesize 'a}\tikzmark{a} \\
            \midrule
            \texttt{\footnotesize arg2} & \texttt{\footnotesize \&'b Bar} & \texttt{\footnotesize None}\\
            \texttt{\footnotesize *arg2} & \texttt{\footnotesize Bar} & \texttt{\footnotesize 'b}\\
            \rowcolor{gray10}
            \texttt{\footnotesize (*arg2).y} & \texttt{\footnotesize String} & \texttt{\footnotesize 'b}\tikzmark{b}\\
            \rowcolor{gray10}
            \tikzmark{c2}\texttt{\footnotesize (*arg2).z} & \texttt{\footnotesize *mut i32} & \texttt{\footnotesize 'b}\tikzmark{c}\\
            \rowcolor{gray10}
            \tikzmark{d2}
            \texttt{\footnotesize *(*arg2).z} & \texttt{\footnotesize i32} & \texttt{\footnotesize 'b}\tikzmark{d}\\
            \midrule
            \texttt{\footnotesize ret} & \texttt{\footnotesize Foo<'a>} & \texttt{\footnotesize None}\\
            \texttt{\footnotesize ret.w} & \texttt{\footnotesize \&'a mut i32} & \texttt{\footnotesize None}\\
            \rowcolor{gray10}
            \tikzmark{e2}
            \texttt{\footnotesize *(ret.w)} & \texttt{\footnotesize i32} & \texttt{\footnotesize 'a}\tikzmark{e}\\
            \texttt{\footnotesize ret.x} & \texttt{\footnotesize *mut String} & \texttt{\footnotesize None}\\
            \rowcolor{gray10}
            \texttt{\footnotesize *(ret.x)} & \texttt{\footnotesize String} & \texttt{\footnotesize 'a}\tikzmark{f}\\
            \bottomrule
        \end{tabular}
        \begin{tikzpicture}[overlay, remember picture, shorten >=.5pt, shorten <=.5pt, transform canvas={yshift=.25\baselineskip, xshift=2pt}]
    \draw [<->] ({pic cs:c}) [bend left] to ({pic cs:e});
    \draw [<->] ([xshift=-11pt]{pic cs:a2}) [bend right] to ([xshift=-2pt]{pic cs:d2});
    \draw [<->] ([xshift=-2pt]{pic cs:d2}) [bend right] to ([xshift=-4pt]{pic cs:e2});
    \draw [line width=0.25mm, red, arrows={<->}] ({pic cs:b}) [bend left] to ({pic cs:f});
    
  \end{tikzpicture}
\end{subfigure}
}
\caption{Motivating example. We calculate borrow lifetimes for each value and look for lifetime bug patterns. The arrows $\longleftrightarrow$ show pairs of values associated with potential violations, and the corresponding rows are 
shaded in {\sethlcolor{gray10}\hl{gray}}. We then use Alias Analysis to filter these violations, leaving only the one shown with a red arrow \red{$\longleftrightarrow$}.}
\label{fig:borrow_lifetime}
\label{fig:bounds}
\end{figure}
\begin{figure}[!t]
    \centering
    \includegraphics[width=\columnwidth]{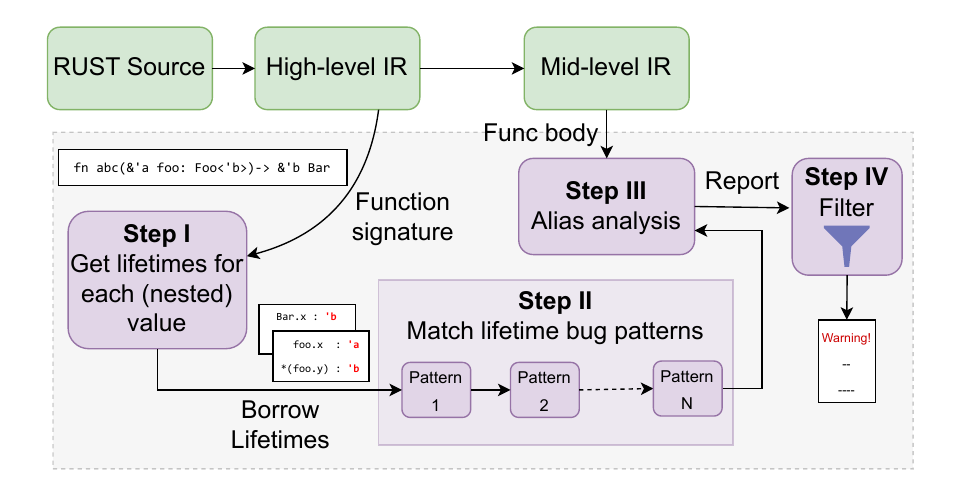}
    \caption{\sysname{} workflow.}
    \label{fig:sys_diagram}
\end{figure}

\noindent
In this subsection, we present \sysname{}, a tool to automatically detect incorrect \las on function signatures in \rust projects. An overview of the entire system is shown in Figure \ref{fig:sys_diagram}. The input to \sysname{} is the source code of a \rust project, and the output is a list of functions with potential \la errors. At a high level, \sysname{} works in four steps.

\begin{itemize}[leftmargin=*]
\item {Step-I.}~\tool extracts lifetimes from each type of a function signature. 
\item {Step-II.}~Next, \tool uses these lifetimes to check for the \la bug patterns identified in~\Cref{sec:patterns} and marks potential buggy functions. 
\item {Step-III.}~The potential buggy functions are then sent to an alias analyzer to check whether values are actually transferred between the locations with \las. 
\item {Step-IV.}~Finally, \tool filters out some common patterns of false positives due to intra-procedural alias analysis. 
\end{itemize}

We now discuss each step of our system in detail and refer~\Cref{fig:borrow_lifetime} as a motivating example to illustrate the steps. In~\Cref{fig:borrow_lifetime}, 
the function signature is: \TT{fn bar<'a,'b>(arg1: \&'a mut i32, arg2: \&'b mut Bar) -> Foo<'a>}. This means that \TT{bar} is a function name, and \TT{'a,'b} are \las associated with the function \TT{bar}. The function takes a parameter \TT{arg1}, which is a mutable reference with borrow lifetime \TT{a} to an \TT{i32} value, and another parameter \TT{arg2}, which is a mutable borrow with lifetime \TT{'b} to a structure of type \TT{Bar}. The function's return type is \TT{Foo<'a>}, which is a structure containing a reference with borrow lifetime \TT{'a}.

\subsubsection{Step-I: Get Borrow Lifetimes}
\label{sec:stageI}

\rebuttal{[Reviewer 1: A reference framework for semantics]}{Our rules for assigning ``outlives'' lifetimes to values are a generalization of the \rust typing rules for borrows, to raw pointers. We propose to treat a raw pointer as though it is ``owned'' by the structure that it is contained in. Each of our rules follows from this assumption. For a formal definition of these rules, please refer to the supplementary material.}

Given a function argument or return type, we extract lifetimes for each nested value as follows:

\be
\item If a value is borrowed with \la \TT{'b}, then the value and all its fields must outlive \TT{'b}, as per Rule1 in~\Cref{sec:back_lifetime}. For example, in~\Cref{fig:borrow_lifetime}, \TT{*arg2} is borrowed with lifetime \TT{'b}, so \TT{*arg2} and \TT{(*arg2).y} must outlive \TT{'b}.

\item If a value is pointed by a raw pointer and the raw pointer is contained inside a structure with \la \TT{'a}, \rebuttal{}{then \TT{'a} is the lifetime of the structure}, and the value must outlive \TT{'a}. For example, in~\Cref{fig:borrow_lifetime}, \TT{ret.x} is a raw pointer inside a structure with \la \TT{'a}, so \TT{*(ret.x)} must outlive \TT{'a}.

\item If a value is pointed by a raw pointer and the raw pointer is contained inside a structure with \textit{no} \la, then
\rebuttal{}{this value must outlive any lifetime that the structure object must outlive.}
For example, in~\Cref{fig:borrow_lifetime}, \TT{(*arg2).z} is a raw pointer in a structure object \rebuttal{}{\TT{*arg2} with no \la. But we derived that \TT{*arg2} must outlive \TT{'b}. So \TT{*(*arg2).z} must also outlive \TT{'b}.}
\ee

\subsubsection{Step-II.~Check for Lifetime Bug Patterns}

After we derive lifetimes for each value, we iterate through these lifetimes and look for potential violations. As we discussed in~\Cref{sec:patterns}, there are two main causes of bugs - assignment from an argument to the returned value (\textbf{arg-return}), or from an argument to another argument (\textbf{arg-arg}).

\noindent
\textbf{Arg-Return:} We iterate through all the values in the arguments, and pair them against all the returned values. We consider only the cases where at least one of the two values is held by a raw pointer, because Rust will automatically infer the lifetimes in all other cases. The patterns that we look out for are:
\be
\item \textit{Use-after-free:} If the types match, and the lifetime of the returned value can be longer than the lifetime of the value in the argument, then this is a potential violation (see Rule1 in~\Cref{sec:back_lifetime}). For example, in~\Cref{fig:borrow_lifetime}, \TT{(*arg2).y} and \TT{*(ret.x)} have the same type \TT{String}, but different lifetimes \TT{'b} and \TT{'a}. This is a potential violation.

\item \textit{Non-exclusive Mutability:} If one value is of type \TT{*mut T} or \TT{\&mut T}, and the other value is of type \TT{T} for some type \TT{T}, then the lifetime of the latter cannot be longer than the lifetime of the former (see Rule2 in~\Cref{sec:back_lifetime}). For example, in~\Cref{fig:borrow_lifetime}, \TT{(*arg2).z} and \TT{*(ret.w)} have types \TT{*mut i32}, and \TT{i32}, but different lifetimes \TT{'b} and \TT{'a}.  This is a potential violation.
\ee

\noindent
\textbf{Arg-Arg:}
We only focus on \textbf{use-after-free} bugs caused by assignments from argument to argument. We take the two arguments at a time, and iterate through all the pairs of values in these arguments. We ignore cases where both values are behind immutable references because an assignment can only be performed if one value is mutable. If the types of the two values match, then we report this as a bug. The actual borrow lifetimes do not matter
, as we discussed in~\Cref{sec:use-after-free}. For example, in~\Cref{fig:borrow_lifetime}, \TT{*arg1} and \TT{*(*arg2).z} have the same type \TT{i32}, and are mutable. So this is a potential violation.

\subsubsection{Step-III.~Perform Alias Analysis}
\label{sec:alias}

The previous steps give us a list of pairs of values such that it would be dangerous to perform an assignment between them. For instance, in~\Cref{fig:borrow_lifetime}, if the value in \TT{(*arg2).y} becomes the value in \TT{*(ret.x)}, then this could lead to a use-after-free. 

This is the problem of alias analysis, or determining whether two pointers point to the same memory location\cite{landi1992undecidability}.
Our algorithm works by computing a ``points-to'' set of values that each pointer may point to. It has the following properties:
\bi
    \item \textit{Intra-procedural:} We perform our analysis on each function body individually, treating function calls as opaque.
    \item \textit{Field-sensitive:} We treat each field of a structure separately.
    \item \textit{Flow-sensitive:} We consider the program statements in order and compute points-to sets at each program point. This is usually infeasible for whole program analysis, but becomes feasible for intra-procedural analysis.
    \ei

We follow a similar setup to Andersen's analysis\cite{andersen1994program}, where we update points-to sets in different ways depending on the type of assignment statement, for example, if we have \TT{a = b} then we merge \TT{b}'s points-to set into \TT{a}'s. However, our algorithm is flow-sensitive because we process each statement just once in order, unlike Andersen's analysis which is flow-insensitive and iterates over the statements till convergence.

\rebuttal{[Reviewer 1: How does alias analysis work in the presence of function calls?]}{For a function call, we simply assume all types of assignment statements between all pairs of function arguments and fields therein. For example, if we have \TT{foo(a, b)}, we separate it into \TT{a = b}, \TT{a = \*b}, \TT{*b = a}, and so on.} We provide pseudocode and more details about our algorithm in the supplementary material.

While implementing field sensitivity, we make a deliberate choice to ignore ``unknown'' fields in function calls. For example, if we have \TT{foo(a, b)}, and \TT{a} has a field \TT{a.x}, then we don't make a connection between \TT{a.x} and \TT{b} unless \TT{a.x} is already ``known'' to our pointer analysis, \ie~it has been used previously. As we will show in~\Cref{sec:design_choices}, we empirically found that this eliminates false positives and increases precision.

\subsubsection{Step-IV.~Apply Shallow Filter}

\rebuttal{[Reviewer 1 and 3: More details about the shallow filter]}{Our shallow filter is designed to filter out false positives, based on some primitive inter-procedural analysis, as well as knowledge of common patterns of false positives. We provide a brief overview here; for more details, please refer to the supplementary material.}

\noindent
\rebuttal{}{
\textbf{\uline{Filter based on \TT{Drop} Implementations:}} 
\noindent
\sysname{} detects violations based on assumptions about how long memory must remain valid. However in certain cases, memory is ``leaked'' and is never freed. This means there is no scope for a use-after-free vulnerability. A memory leak is \textbf{not} considered an exploitable vulnerability or undefined behavior in \rust.
}

\noindent
\rebuttal{}{
Hence we implement a simple inter-procedural check. If the value being re-assigned is behind a raw pointer in \textbf{both} the source as well as the target structures, then we only mark this as a violation if \textbf{either} of those structures has a custom \TT{Drop} implementation. A custom \TT{Drop} implementation would usually involve cleaning up memory, and this means that when that structures gets dropped, the \TT{data} value might be freed.
}

\noindent
\rebuttal{}{
\textbf{\uline{Filter Common Patterns of False Positives:}} There are certain common traits in \rust that rely on the semantic correctness of their implementation to ensure their safety. One such example is the \TT{IteratorMut} trait, specifically its \TT{next} and \TT{next\_back} methods \cite{splitting_borrows}.
}

\noindent
\rebuttal{}{
There are three specific functions from trait implementations that we filter out - \TT{Iterator::next}, \TT{Iterator::next\_back}, \TT{Clone::clone}. It is true that there can be potential flaws in their implementations, which is why the shallow filter is the last stage of our analysis, and a user has the option of turning it off.
}
\section{Experimental Setup}
\label{sec:setup}


\subsection{Implementation}

When \rust code is compiled, it goes through several intermediate representations (IRs), two of which are the High-level IR (HIR) and the Mid-level IR (MIR). The HIR is a tree representation of the program that is similar to an Abstract Syntax Tree (AST). The MIR represents each function as a control-flow graph of basic blocks, with each basic block consisting of a number of statements. MIR boils the different kind of statements occurring in a program down to a core set of statement types like assignments, memory allocation, and memory de-allocation. We use a hybrid approach that combines information from the HIR and the MIR. We use the HIR to get struct definitions and function signatures, and use the MIR to perform alias analysis.

We implement our system \sysname{} in \rust as a fork of the Rudra \cite{bae2021rudra} project. We use the \TT{nightly-2022-11-18} version of \TT{rustc} and \TT{cargo}, and run our tool on \rust projects as a \TT{cargo} subcommand.
\subsection{Study Subject}
We evaluate \sysname{} in three different settings: (i) \rebuttal{}{23} real-world  vulnerable functions curated from 9 crate libraries, (ii) a synthetic vulnerability dataset, and (iii) running \tool across 372 popular Rust crate libraries. 
This section describes each setting in detail. 

\subsubsection{Curating bugs from Vulnerability Reports.}
\label{sec:rustsec}

A library crate (or \textit{crate} for short) in Rust is a project that provides functionality that can be used in other Rust projects.
This is similar to a package or library in other programming languages. Rust hosts a community crate registry at \texttt{crates.io}, and any project can add a library crate from this registry as its dependency. 
Adding a crate as a project dependency allows one to use APIs from that crate in the project.
The RustSec Security Vulnerability Database\cite{rustsec} is a collection of security vulnerability reports in Rust crates. We manually inspected each report in the RustSec database \cite{rustsec}, looking for issues related to incorrect \la. 
We found 9 vulnerabilities that matched this pattern. These are shown in Table \ref{tab:real_world}. Each of these corresponds to one crate containing one or more buggy functions. There are a total of \rebuttal{}{23} functions with bugs across the 9 vulnerabilities.

\subsubsection{Creating a Synthetic Bug Dataset.}
\label{sec:synthetic}

In addition to evaluating \sysname{} on the bugs from vulnerability reports, we would like to check whether it can generalize to similar patterns of bugs. To do this, we prepared a synthetic bug dataset based on the above 9 vulnerability reports in three steps, as follows.
\bi
\item \textit{Program Slicing and Simplification.} Each of the 9 reports contains one or more functions with a \la 
bug. We selected one function from each report, isolated the structures and functions responsible for the bug, and rewrote them in a minimal form. 
This gave us a list of 9 synthetic bugs.
\item \textit{Program Transformations.} For each of the 9 synthetic bugs from the previous step, we \rebuttal{[Reviewer 2: is this [synthetic dataset] manually generated?]}{manually} applied a variety of transformations to them that retain the semantics of the code but change the high level structure. For instance, we wrap borrows in structures or enums, change concrete types into generic types, change datatypes, add more dummy borrows with different lifetime annotations, etc. We transformed the 9 bugs in two different ways each by applying selected transformations, to get a total of 27 synthetic bugs.
\item \textit{Exploit Construction.} For each synthetic bug, we construct an ``exploit'' function that showcases the vulnerability present in it and triggers undefined behavior.
\ei


\subsubsection{Scanning Crates from the Registry}
\label{sec:registry}

We also use our tool to scan crates from the crates.io registry. We pick the top 2000 most downloaded crates as of November 15th, 2022, and filter out those which do not use unsafe code or lifetime annotations. This leaves us with 372 crates. We download the original source for each of them using the \TT{cargo-download} tool. Our tool is implemented as a \TT{cargo} subcommand, and we run this subcommand on the downloaded source folders.

\subsection{Evaluation and Metrics}

To evaluate \tool, we first define  a ``ground truth'' bug. 
For the RustSec and Synthetic bug dataset, we know by construction that each of the functions reported in the security vulnerability are confirmed bugs.
For all \textit{other} functions that our tool marks as potential bugs, we manually inspect each one. We classify a reported bug as genuine only if we are able to clearly exploit the vulnerability to cause undefined behavior. If we are uncertain or unable to cause undefined behavior, we classify this as a false positive.

We assess \sysname{}'s ability to identify lifetime annotation bugs using precision and recall metrics. 
In a given dataset, if the bugs detected by \sysname{} are denoted by $S_{\sysname{}}$, and the ground truth bugs are denoted by $S_{GT}$, the precision and recall of \sysname{} can be defined as 
$|S_{GT} \cap S_{\sysname{}}|/|S_{\sysname{}}|$ and $|S_{GT} \cap S_{\sysname{}}|/|S_{GT}|$
respectively.


\subsection{Baselines}
We evaluate against the following baselines:
\bi
\item \textbf{Rudra\cite{bae2021rudra}:} Rudra is a static analyis based bug detector that detects certain kinds of errors in unsafe \rust.
\item \textbf{MirChecker\cite{li2021mirchecker}:} MirChecker is a static analysis based analyzer that can detect certain categories of undefined behavior, like overflows and out-of-bounds accesses.
\item \textbf{Miri\cite{miri}:} Miri is a dynamic analysis tool that interprets one MIR instruction at a time and detects memory errors. 
\ei
\section{Research Questions}
\label{sec:rqs}

We aim to answer the following research questions:
\bi
\item \textbf{RQ1: Effectiveness of our tool.} We evaluate the effectiveness of \sysname{} in three settings: 1) Can \sysname{} detect reported \rust security vulnerabilities that result from incorrect \las? 
2) How does \sysname{} generalize to the synthetic bug dataset that we curated? 
3) How does \sysname{} perform when we run it on a large set of real-world \rust crates?

\item \textbf{RQ2: Impact of Design Choices.} How do the different model components of \sysname{} impact its performance on vulnerabilities and synthesized bugs?

\item \textbf{RQ3: Existing tools.} Are there any existing tools that can detect incorrect lifetime annotations?
\ei

\subsection{RQ1: Effectiveness of our Tool}
\label{sec:rq_rustsec}

This research question is structured as follows. First, we shall evaluate the performance of \sysname{} by running it on the \rebuttal{}{23} functions from 9 crates from RustSec vulnerabilities. Then, we will measure the ability of \sysname{} to generalize by running it on our synthetic bug dataset of 27 bugs. Finally, we run \sysname{} on a large set of real-world \rust crates.

\begin{table}
    \caption{\small{Running \tool on Rustsec vulnerabilities due to \la bugs. FN=False Negatives, FP=False Positives}}
    \label{tab:rustsec}
    \centering
    \begin{tabular}{c|c|c|c|c|c}
        \toprule
        & \textbf{Error} & \textbf{\#Vulnerable} &  &  & \\
        \textbf{RustSec ID} & \textbf{Type}  & \textbf{Funcs} & \textbf{\#Detected} & \textbf{\#FP} & \textbf{ \#FN}\\
        \toprule
        2018-0020 & {UAF} & 1\rebuttal{}{\ + 2}\red{*} & 2 & \rebuttal{}{0} & 1\\
         
        2020-0023 & {NEM} & 2 & 2 & 0 & 0\\
         
        2020-0060 & {UAF} & 1 & 0 & 0 & 1\\
         
        2021-0128 & {UAF} & 7 & 7 & 2 & 2\\
         
        2021-0130 & {UAF} & 2 + 3\red{*} & 5 & 0 & 0\\
         
        2022-0028 & {UAF} & 2 & 0 & 0 & 2\\
         
        2022-0040 & {UAF} & 1 & 0 & 0 & 1\\
         
        2022-0070 & {UAF} & 1 & 0 & 0 & 1\\
        
        2022-0078 & {UAF} & 1 & 0 & 0 & 1\\
        \midrule
        \textbf{Total} & & \rebuttal{}{\textbf{23}} & \textbf{16} & \rebuttal{}{\textbf{2}} & \textbf{9}\\
        \bottomrule
    \end{tabular}
    \vspace{4pt}
    \\
    \footnotesize{
    The asterisk \red{*} denotes newly exposed bugs that weren't part of the initial report.}
\end{table}

\subsubsection{Evaluation on RustSec Vulnerabilities} We run \sysname{} on the 9 crates from RustSec vulnerabilities, containing 21 buggy functions. Its performance is summarized in~\Cref{tab:rustsec}. Overall, the precision is \rebuttal{}{\textbf{87.5\%}}, and the recall is \rebuttal{}{\textbf{60.9\%}}.

\noindent
\uline{\textbf{Detected bugs:}} \sysname{} generates 16 vulnerability reports from 3 crates. 2 of these reports are NEM bugs, and the other 14 are UAF. The 3 bugs in~\Cref{tab:ex_bugs} are among the vulnerable functions that \sysname{} is able to detect correctly.


\noindent
\uline{\textbf{New bugs exposed:}}
For \TT{RUSTSEC-2021-0130} corresponding to the \TT{lru-rs} crate, the report had only 2 functions listed as bugs. However, our tool detected 3 \textit{additional} functions as potential bugs - \TT{peek\_lru}, \TT{get\_or\_insert}, \TT{get\_or\_insert\_mut}. When we looked at the commit history on GitHub\cite{lru}, it turned out that these functions had been subsequently identified as memory safety bugs.

\noindent
\rebuttal{}{
In the \TT{pulse-binding} crate corresponding to the vulnerability RUSTSEC-2018-0020, \sysname{} flags two additional functions as potential bugs. We have corresponded with the developer\cite{pulse_binding} regarding these functions, and these seem to be genuine vulnerabilities that can be exploited with appropriate driver code to cause runtime errors with Valgrind.}

\noindent
The fact that \sysname{} was able to detect these bugs despite them not being explicitly labeled is a strong validation of our tool and approach.

\noindent
\uline{\textbf{False Positives:}} \rebuttal{}{In the \TT{rusqlite} crate, \sysname{} flags two additional functions as potential bugs, however these are not exploitable vulnerabilities. Listing \ref{lst:interrupt} shows one of them:}
\lstset{xleftmargin=0.07\columnwidth, xrightmargin=0.05\columnwidth, escapeinside={<@}{@>}}
\begin{lstlisting}[caption=False positive in the \TT{rusqlite} crate corresponding to \TT{RUSTSEC-2021-0128}, captionpos=b, label=lst:interrupt, language=Rust, frame=none, style=colouredRust, basicstyle=\scriptsize \ttfamily, numbers=left]
pub struct InnerConnection {
    pub db: *mut ffi::sqlite3,
    interrupt_lock: Arc<Mutex<*mut ffi::sqlite3>>, ...
}
impl InnerConnection {
    #[inline]
    pub fn get_interrupt_handle(&self)->InterruptHandle
    {   InterruptHandle {
            db_lock: Arc::clone(&self.interrupt_lock),
        }
    }
    pub fn close(&mut self) -> Result<()> {
        ...
        let mut shared_handle = self.interrupt_lock
                                    .lock().unwrap();
        ...
            *shared_handle = ptr::null_mut();
            ...
    }
}
impl InterruptHandle {
    pub fn interrupt(&self) {
        let db_handle = self.db_lock.lock().unwrap();
        if !db_handle.is_null() {
            unsafe{ffi::sqlite3_interrupt(*db_handle)}
        }
    }
}
\end{lstlisting}
\rebuttal{}{
The pointer \TT{interrupt\_lock} is shared between the structure \TT{InnerConnection} and \TT{InterruptHandle} on line 9, and there is no \la on \TT{InterruptHandle} to denote this. But the implementations of the \TT{close} and \TT{interrupt} functions circumvent any vulnerability. The code sets the pointer to null on Line 17, and checks if it is null before dereferencing it on line 24. This example illustrates how the difference between safe code and a vulnerability can be very subtle. At the same time, because this is so subtle, this indicates a risk in the future development of this crate and can be termed a ``code smell''. We discuss more about these patterns in \Cref{sec:real_world}.
}

\noindent
\uline{\textbf{False Negatives:}} Our tool is unable to detect 9 of the \rebuttal{23} functions. Some of the reasons are:
\be
\item \textit{``Hidden'' Raw Pointers:} Our tool reports violations only in cases where there is a raw pointer involved. Sometimes, it is not easy to infer that a certain type contains a raw pointer. 
\TT{pub fn waker<W>(wake: Arc<W>) -> Waker \{ ... \}} is an example. 

Here, \TT{Waker} is a structure containing a raw pointer, but we don't have access to the definition of this structure because it is an external dependency. So our tool ignores this. \textbf{4} out of the \textbf{9} false negatives are because of hidden raw pointers. \rebuttal{[Reviewer 1: Fetching definitions from other crates]}{We discuss more about this limitation in \Cref{sec:threats}}.

\item \textit{Type conversion:} To increase precision, \tool checks if the types of the source and target values match. But that can also be too restrictive, as Listing \ref{lst:aggregate} shows:
\lstset{xleftmargin=0.08\columnwidth, xrightmargin=0.05\columnwidth, escapeinside={<@}{@>}}
\begin{lstlisting}[caption=False negative in the \TT{rusqlite} crate corresponding to \TT{RUSTSEC-2021-0128}, captionpos=b, label=lst:aggregate, language=Rust, frame=none, style=colouredRust, basicstyle=\footnotesize \ttfamily, numbers=left]
pub struct InnerConnection {
    pub db: *mut ffi::sqlite3, ...
}
pub struct Connection {
    db: RefCell<InnerConnection>, ...
}
impl Connection {
    pub fn create_aggregate_function<A, D, T>(
        &self, ...,
        aggr: D,
    ) -> Result<()>
    where
        A: RefUnwindSafe + UnwindSafe,
        D: Aggregate<A, T>,
    ...
\end{lstlisting}
Here, the type of the raw \TT{db} pointer on line 2 is \TT{ffi::sqlite3}, and the type of \TT{aggr} on line \rebuttal{10} is a generic type \TT{D}. \tool would filter this out because the types do not match, but in fact, \TT{D} is converted to a \TT{c\_void} pointer which is then attached to the \TT{db} using a C foreign function call (not shown in the snippet). This kind of type conversion is difficult to reason about.
\textbf{2} of the \textbf{9} false negatives are due to type conversion.

\noindent
\rebuttal{[Reviewer 3: Causes of the remaining 3 false negatives]}{Of the remaining 3, one is due to \las on the arguments of a closure, which is something that we do not reason about. One is due to a shortcoming of alias analysis, and the last one is due to a complex \textit{indirect} lifetime semantics where one value acts as a drop guard for \textit{another} value.}

\ee



\subsubsection{Evaluating Generalizability using Synthetic Dataset.}

We run \sysname{} on the 27 synthetic bugs curated. For each of the 9 original RustSec vulnerability reports, we derive 3 synthetic bugs. The three synthetic bugs derived from the n-th RustSec report are denoted by \TT{n.1}, \TT{n.2} and \TT{n.3}.

~\Cref{tab:synthetic} summarizes the result---\sysname{} is able to detect 16 out of 27 bugs, corresponding to a recall of \textbf{59.3\%}. There are some cases where \sysname{} is unable to detect the bug in the original RustSec report, but is able to detect one or more of the corresponding bugs in the synthetic dataset. A few of the reasons for this are as follows:
\bi
\item \textit{We re-implement some external dependencies}. For example, 
\sysname{} is unable to detect the bug in \TT{pub fn waker<W>(wake: Arc<W>) -> Waker}, because it does not have access to the implementation of the \TT{Waker} structure. But in our synthetic examples 3.1 and 3.2, we re-implement the \TT{Waker} structure, and thus \sysname{} is able to see that there is a raw pointer inside.

\item \textit{Some types are simplified.} For example in the 6th RustSec report \TT{RUSTSEC-2022-0028}, \sysname{} is unable to detect a raw pointer because it is ``hidden'' behind a chain of type re-definitions and imports. In our synthetic examples 6.1 and 6.2, the types are simplified and it detects the raw pointer.
\ei


Since we constructed this dataset ourselves, there aren't many functions apart those containing bugs. So our tool does not report any false positives on this dataset.

\begin{table}
    \centering
    \small
    \caption{Results of \sysname{} on synthetic bug dataset.}

    \label{tab:synthetic}
    \resizebox{0.8\columnwidth}{!}{%
    \begin{tabular}{cc|cc|cc}
    \toprule
    Bug & Detected & Bug & Detected & Bug & Detected \\
    \midrule
    1.1 & No & 4.1 & Yes & 7.1 & No \\
    1.2 & No & 4.2 & Yes & 7.2 & No \\
    1.3 & No & 4.3 & No & 7.3 & No \\
    \midrule
    2.1 & Yes & 5.1 & Yes & 8.1 & No \\
    2.2 & Yes & 5.2 & Yes & 8.2 & Yes \\
    2.3 & Yes & 5.3 & Yes & 8.3 & Yes \\
    \midrule
    3.1 & Yes & 6.1 & Yes & 9.1 & Yes \\
    3.2 & Yes & 6.2 & Yes & 9.2 & Yes \\
    3.3 & No & 6.3 & No & 9.3 & No \\
    \bottomrule
    \end{tabular}
    }
\end{table}

\subsubsection{Evaluation on Real World \rust Crates.}
\label{sec:real_world}

\rebuttal{[Reviewer 1, 2 and 3: Evaluation on real-world crates needs to be more thorough.]}{} We would like \sysname{} to be able to quickly scan and flag potential bugs in Rust crates, and scale to the thousands of crates in the Rust crate ecosystem. We run \sysname{} on a filtered list of 372 Rust crates as described in Section \ref{sec:registry}. \sysname{} is able to scan these crates in just under 2 hours, and flags \rebuttal{}{\textbf{85}} potential bugs from \rebuttal{}{\textbf{27}} crates.

\noindent
\rebuttal{}{
Each of these bugs was classified into two broad categories - bugs which can be exploited to cause undefined behavior, and ``code smells'', \ie code patterns that cannot be exploited, but could indicate potential for a vulnerability to arise during future development. We have further grouped the code smells into different patterns. The results are summarized in \Cref{tab:real_world}. Here we discuss some of these patterns. For a complete discussion with examples, please refer to the supplementary material.
}

\begin{table}
    \centering
    \small
    \caption{Running \sysname{} on real-world \rust crates}
    \resizebox{0.8\columnwidth}{!}{%

    \begin{tabular}{c|c|c}
        \toprule
         \textbf{Category} & \textbf{Sub-category} & \textbf{\#}\\
         \midrule
         Exploitable bugs & & \textbf{3}\\
         \midrule
         \multirow{ 5}{*}{Code smells} & Freed ptr but no deref  & 14\\
            & User-implemented ref counting & 30\\
            & Memory copy, not alias & 21\\
            & Different field of struct & 16\\
            & Two lifetime annotations & 1\\
         \midrule
            \textbf{Total} & & \textbf{85} \\
        \bottomrule
    \end{tabular}
    }
    \label{tab:real_world}
\end{table}

\noindent
\rebuttal{}{\textbf{\uline{Exploitable Bugs:}} \sysname{} finds \textbf{3 new bugs} among these crates that can be exploited to cause undefined behavior, despite the fact that these are within the top $\sim$1.5\% most popular crates, with tens of millions of downloads each. This is in addition to the \textbf{2 new bugs} from \TT{pulse-binding} described earlier.} Listing \ref{lst:cslice} shows one from \TT{cslice}:

\lstset{xleftmargin=0.08\columnwidth, xrightmargin=0.05\columnwidth, escapeinside={<@}{@>}}
\begin{lstlisting}[caption=A genuine bug that \sysname{} found in \TT{cslice}, captionpos=b, label=lst:cslice, language=Rust, frame=none, style=colouredRust, basicstyle=\footnotesize \ttfamily, numbers=left]
pub struct CMutSlice<'a, T> {
  base: *mut T, ...
}
pub fn as_mut_slice(&mut self) -> &'a mut [T]
{   unsafe { slice::from_raw_parts_mut(
            self.base, self.len)} }
\end{lstlisting}
On line 4, the lifetime \TT{'a} doesn't match with the lifetime of the borrow \TT{\&self} which has an implicit anonymous lifetime. So this allows aliased mutable references.


\noindent
\rebuttal{}{
\uline{\textbf{Freed Pointer but no Deref:}} There is a pointer within a structure that is freed, but all the methods that dereference the pointer are either private or \TT{unsafe}. A small modification like making the method public or removing the unsafe label could result in an exploitable vulnerability.
}

\noindent
\rebuttal{}{
\uline{\textbf{User-implemented Reference Counting:}} The developer has implemented their own system of reference counting and memory management. Every time a pointer is duplicated or dropped, an internal reference count within the memory location is incremented or decremented. The memory is freed only if the count has reached zero. If any of these increments, decrements or checks are omitted, this could result in a UAF or memory leak.
}

\noindent
\rebuttal{}{
\uline{\textbf{Memory copy, not alias:}} \sysname{} predicts that there could be a vulnerability if two pointers were \textit{aliased}, however the function actually \textit{copies} the underlying memory of the first pointer to create the target of the second pointer.
}

\noindent
\rebuttal{}{
\uline{\textbf{Different field of structure:}} \sysname{} predicts that there could be a vulnerability if two pointers were aliased, however the aliasing actually happens with a \textit{different field} of a structure.
}

\noindent
\rebuttal{}{
\uline{\textbf{Two lifetimes:}} The structure has \textit{two} \las, say \TT{'a} and \TT{'b}, corresponding to two raw pointers within the structure. One of the raw pointers is returned as a borrow with annotation \TT{'a}. But this is only valid if the underlying memory actually outlives \TT{'a}, and not \TT{'b}. This relies on the semantics of the functions that allocate memory for these pointers.
}



\begin{myshadowbox}
\textbf{Summary:} We evaluate \sysname{} in different settings. On the RustSec vulnerabilities, it achieves a precision of \rebuttal{}{\textbf{87.5\%}} and a recall of \rebuttal{}{\textbf{60.9\%}}. On our synthetic dataset, it achieves a recall of \textbf{59.3\%}. Finally, we were able to use \sysname{} to scan 357 \rust crates ``in the wild'' in under 2 hours. \rebuttal{}{We found 3 exploitable bugs and 82 code smell patterns.}
\end{myshadowbox}

\subsection{RQ2: Impact of Design Choices}
\label{sec:design_choices}

We investigate the effect of three design choices in \sysname{} - the Alias Analysis module, our handling of unknown fields in Alias Analysis, and the Shallow Filter.

\subsubsection{Alias Analysis.}
We investigate the effect of the alias analysis module by removing it entirely.  
As shown in in~\Cref{tab:unknown_fields}, the recall increases, but the precision drops tremendously from \rebuttal{}{87.5\% to 34.1\%}. This shows that simple pattern matching is not sufficient, and deeper pointer analysis is necessary.

\subsubsection{Unknown Fields in Alias Analysis.}
In~\Cref{sec:alias}, we described how we made a choice to exclude ``unknown'' fields in function calls, while designing our Alias Analysis algorithm. Here we explore the impact of that choice.
The results are shown in~\Cref{tab:unknown_fields}. We can see that considering unknown fields increases the recall slightly,
but the precision on RustSec falls from \rebuttal{}{87.5\% to 46.9\%}.

\begin{table}[h]
    \small
    \centering
     \caption{Impacts of different design choices. The numbers that are different from the original are shown in bold.}
    \label{tab:unknown_fields}
    \rebuttal{}{
    \begin{tabular}{c|c|c|c|c}
    \toprule
            & \multicolumn{2}{c|}{RustSec} & \multicolumn{2}{c}{Synthetic} \\
            \midrule
                & Recall & Prec. & Recall & Prec.\\
            \midrule
        Original (\sysname{}) &  60.9\%  &  87.5\%  &  59.3\%  & \multirow{4}{*}{--} \\
        w/o Shallow Filter & 60.9\% & \textbf{63.6\%} & 59.3\% &  \\
        w/ Unknown Fields &  \textbf{65.2\%}  &  \textbf{46.9\%}  &  \textbf{62.9\%}  &  \\
        w/o Alias Analysis & \textbf{65.2\%} & \textbf{34.1\%} & \textbf{74.1\%} & \\
        \bottomrule
    \end{tabular}
    }
\end{table}

\subsubsection{Shallow Filter.}
Our shallow filter is used to filter out common patterns of false positives that result from trait implementations. We run \sysname{} on the \rebuttal{}{21} functions from RustSec reports. With the filter, there were \rebuttal{}{2} false positives; without it, that number increases to \rebuttal{}{\textbf{8}}. The precision drops from \rebuttal{}{\textbf{87.5\%} to \textbf{63.6\%}}.

\begin{myshadowbox}
\textbf{Summary:} All three of the design choices we investigate contribute significantly to the precision of \sysname{}.
\end{myshadowbox}

\subsection{RQ3: \minorrevision{}{Comparison with} existing tools}
\label{sec:baselines}

In this section, we aim to evaluate existing static analysis tools for detecting memory safety bugs in \rust. 
\rebuttal{[Reviewer 3: No apples-to-apples comparison in RQ3]}{Although none of these tools are designed specifically to detect \la bugs, we investigate if there is any overlap between \la bugs and the patterns that these tools look for. This also highlights the point that \la bugs are not just generic memory safety bugs and require special treatment.}

\begin{table}
    \centering
    \minorrevision{}{
    \resizebox{0.9\columnwidth}{!}{%
    \begin{tabular}{c|c|c|c}
        \toprule
         \textbf{Tool} & \# Bugs & Tool detected & \sysname{} detected\\
         \midrule
         Rudra & 50 & 0 & 30 \\
         MirChecker & 27 & 0 & 16 \\
         Miri (w/o exploit code) & 27 & 0 & 16 \\
         Miri (w. exploit code) & 27 & 27 & 16 \\
         \bottomrule
    \end{tabular}
    }
    \caption{\rebuttal{}{Comparison with existing tools}}
    }
    \label{tab:baselines}
    \vspace{-4mm}
\end{table}

\subsubsection{Rudra\cite{bae2021rudra}:}\minorrevision{[Reviewer 2: RQ3 is not substantial enough]}{Rudra is a static analysis tool that identifies three types of potential memory and thread safety violations - panic safety, higher order safety invariant, and Send/Sync variance. It is not explicitly designed to detect \la bugs; however, we would like to check if there is any overlap between Rudra's bug categories and \la bugs.}

\noindent
We run Rudra on the \rebuttal{}{23} RustSec bugs and our synthetic bug dataset of 27 bugs. We found that Rudra was unable to detect \textit{any} of these \rebuttal{}{50} bugs. \minorrevision{}{This indicates that there is no overlap between Rudra's bug patterns and \la bugs.}.

\subsubsection{MirChecker\cite{li2021mirchecker}:}\minorrevision{}{MirChecker is a static analysis tool for the Rust MIR based on the theory of Abstract Interpretation. It forms a control-flow graph (CFG) of the program, defines a transfer function for each value and statement, and propagates these values across the CFG to reach a fixed point. It then uses these values to detect potential vulnerabilities in Rust code.}

This project is old and not compatible with the toolchains used by most current Rust projects, but we were able to run it on our synthetic bug dataset. MirChecker was unable to detect \textit{any} of the 27 bugs. 
This shows that 
MirChecker will also not be able to detect other more complicated bugs.

\subsubsection{Miri\cite{miri}:}\minorrevision{}{Miri is an interpreter for the Rust MIR. It emulates the execution of each line of code using a virtual address space, and can detect memory leaks, out of bounds accesses, and other potential vulnerabilities. This is a \textit{dynamic} tool, and deals with actual vulnerabilities that arise when Rust code is run. In contrast, \sysname{} deals with \textit{potential} vulnerabilities that \textit{could} arise when a function is called in a certain context. So we expect Miri to detect \la bugs if and only if the exploit code is also provided.}

For each of the 27 synthetic bugs, we ran Miri directly with the exploit function (see~\Cref{sec:synthetic}) that exposes the vulnerability. Miri detected a memory safety error in all 27 of the cases, but without the exploit function, Miri cannot detect any one of them. This shows that given a large enough test suite, Miri might be able to find these bugs, but constructing that test suite is non-trivial.
We discuss more about dynamic tools in~\Cref{sec:related_work}.

\begin{myshadowbox}
\textbf{Summary:} Existing static analysis tools are not well-suited to this problem. For instance, Rudra and MirChecker were \textbf{unable to detect any} of the 
lifetime annotation bugs. The dynamic tool Miri, in contrast, was able to detect the bugs, but it requires manually crafted test cases, which is non-trivial in real project settings. 
\end{myshadowbox}
\section{Threats to Validity}
\label{sec:threats}


Our dataset from vulnerability reports includes only 23 functions from 9 crates, potentially limiting generalizability. To address this, we synthesized 27 vulnerable functions with varied contexts and types. Although the synthetic data involves some subjective decisions, the dataset is publicly available for inspection. Additionally, we analyzed \sysname{} on several popular Rust crates, evaluating a total of 50 vulnerable functions across more than 300 real Rust libraries.


On the dataset of functions from vulnerability reports, \sysname{} achieves a precision of 87.5\% and a recall of 60.9\%. There is a tradeoff between these metrics, as higher recall can be achieved at the expense of precision, as shown in \ref{sec:design_choices}. Note that, this is a challenging problem, and \textit{no} other tool can handle such bugs. We believe our precision is comparable to or better than other Rust static analysis tools; for example, Rudra has a precision of 11.05\%, and MirChecker has 4.9\%.

\minorrevision{[Reviewer 1: Fetching definitions from other crates]}{Our implementation uses the HIR for identifying error patterns and the MIR for alias analysis. However, the HIR does not contain definitions of structures from external crates and dependencies. This causes a loss of precision or recall. In theory, one could re-implement our tool entirely at the MIR level and have access to all external dependencies; however this would come with its own set of challenges since a lot of the source information is lost in the MIR. We plan to address this in future work.}
 
\minorrevision{[Reviewer 1: System seems ad-hoc]}{More broadly speaking, \sysname{} is a tool with a single specific use-case, \ie, detecting bugs in unsafe Rust arising due to lifetime annotation errors. It does not generalize to, or provide insights for, other kinds of unsafe Rust analysis.}


\section{Related Work}
\label{sec:related_work}


\textbf{Dynamic testing of \rust code:} Miri \cite{miri} is an interpreter 
to detect invalid memory accesses or memory leaks (see ~\Cref{sec:baselines}). Stacked Borrows\cite{jung2019stacked} (incorporated into Miri) develops a ``stack'' model to reason about borrows \textit{without} explicitly using lifetimes. Fuzzing with an address sanitizer is another way to detect potential errors, but existing \rust fuzzers~\cite{fuzzcheck, cargofuzz} are not well suited to our problem. \TT{cargo-fuzz} \cite{cargofuzz} is a fuzzing tool for \rust that uses \TT{libFuzzer} \cite{libfuzzer} as a backend. However, it is restricted to fuzzing functions that only accept input strings of a fixed format and type. 


\textbf{Static analysis for detecting bugs in \rust:}
MirChecker \cite{li2021mirchecker} and SafeDrop \cite{cui2021safedrop} are static analyzers on the \rust MIR that detect certain memory safety errors. However, these approaches tend to have 
high false negatives. 
Rudra \cite{bae2021rudra} identifies three categories of bugs arising from unsafe code, and scans the entire \rust crate ecosystem for these bug patterns. Rudra is the closest work to ours in the sense that it uses static analysis to identify potential memory safety violations. However, Rudra can handle only three specific kinds of errors, and cannot detect memory safety bugs caused by incorrect lifetime annotations.

\textbf{Empirical analysis of unsafe \rust:} Recent works~\cite{qin2020understanding, evans2020rust, astrauskas2020programmers, xu2020memory} characterize how developers use unsafe code in \rust, and build an empirical understanding of bugs. 

\textbf{Formal semantics of \rust:} Patina \cite{reed2015patina} 
formalizes the type system of \rust. FR \cite{pearce2021lightweight} is a lightweight formalism of \rust that captures type checking and borrow checking. Oxide \cite{weiss2019oxide} models \rust code with non-lexical lifetimes \cite{nonlexicallifetimes} and a syntax close to \textit{safe} surface \rust. However,
these methods do not model unsafe code. RustBelt \cite{jung2017rustbelt} develops a comprehensive formal semantics based on lifetimes for a realistic subset of \rust including unsafe code.
However, this cannot scale to larger codebases and crates. Verifying the safety of a given library with RustBelt involves writing proofs with the Coq proof assistant \cite{coqproofassistant}, which requires considerable expertise and effort.




\section{Conclusion}
\label{sec:conclusion}

In this paper, we investigate memory safety bugs arising from incorrect lifetime annotations in the \rust language. We demonstrate how such bugs can arise, and characterize bug patterns. Leveraging these patterns, we design \sysname{} to detect incorrect lifetime annotations in \rust. We show that \sysname{} is able to effectively detect bugs in real-world \rust projects, compared to existing approaches which are unable to detect \textit{any} of these bugs. 
Future work could involve making the static analysis framework more precise, and including other kinds of bugs in unsafe \rust.

\bibliographystyle{IEEEtran}
\bibliography{main}


\pagebreak
\clearpage
\appendices
\raggedcolumns
\section{Bug Reports from Crates}

In \Cref{sec:real_world}, we presented the results of running \sysname{} on some of the most frequently downloaded \rust crates. Here we discuss these results in more detail, with examples:

\subsection{Exploitable Bugs:}
\sysname{} finds 3 exploitable bugs, 2 from \TT{cslice}, and 1 from \TT{bv}. We presented one from \TT{cslice} in the main text; here are both. We have reported these bugs to the developer\footnote{\url{https://github.com/dherman/cslice/issues/5}}.

\lstset{xleftmargin=0.08\columnwidth, xrightmargin=0.05\columnwidth, escapeinside={<@}{@>}}
\begin{lstlisting}[caption=Two exploitable bugs in \TT{cslice}, captionpos=b,, language=Rust, frame=none, style=colouredRust, basicstyle=\footnotesize \ttfamily, numbers=left]
pub struct CMutSlice<'a, T> {
  base: *mut T, ...
}
pub fn as_slice(&mut self) -> &'a mut [T]
{   unsafe { slice::from_raw_parts(
            self.base, self.len)} }
            
pub fn as_mut_slice(&self) -> &'a [T]
{   unsafe { slice::from_raw_parts_mut(
            self.base, self.len)} }
\end{lstlisting}

Here is one from \TT{bv}, that we have reported to the developer\footnote{\url{https://github.com/tov/bv-rs/issues/16}}.

\lstset{xleftmargin=0.08\columnwidth, xrightmargin=0.05\columnwidth, escapeinside={<@}{@>}}
\begin{lstlisting}[caption=An exploitable bug in \TT{bv}, captionpos=b, language=Rust, frame=none, style=colouredRust, basicstyle=\scriptsize \ttfamily, numbers=left]
impl<'a, Block: BlockType> BitSliceMut<'a, Block> {
  /// Creates a `BitSliceMut` from a mutable 
  /// array slice of blocks.
  /// ...
  pub fn from_slice(blocks: &mut [Block]) -> Self {
    BitSliceMut {
      bits:   blocks.as_mut_ptr(),
      span:   SliceSpan::from_block_len::<Block>
                                  (blocks.len()),
      marker: PhantomData,
    }
  }
  ...
}
\end{lstlisting}

This creates a mutable pointer into the slice, but the slice itself can be dropped while the pointer is still accessible. The lifetime annotation \TT{'a} on line 1 is unrelated to the anonymous annotation \TT{'\_} on line 5.

In both the issues that we have filed, we have included code that exploits the vulnerability, verified with Miri\cite{miri}.

\subsection{Freed Pointer but no Deref}
These are cases where there is a pointer within a structure that is freed, but all the methods that dereference the pointer are either private or unsafe. A small modification like removing the unsafe label or making the method private could result in an exploitable vulnerability. This is one such example from \TT{wasmi}.
\lstset{xleftmargin=0.08\columnwidth, xrightmargin=0.05\columnwidth, escapeinside={<@}{@>}}
\begin{lstlisting}[caption=A code smell pattern in \TT{wasmi}, captionpos=b, language=Rust, frame=none, style=colouredRust, basicstyle=\scriptsize \ttfamily, numbers=left]
pub struct ValueStack {
    values: Vec<UntypedValue>,
    sp: usize,
    max_sp: usize,
}
pub struct FrameRegisters {
    ptr: *mut UntypedValue,
}
impl ValueStack {
    pub fn root_stack_ptr(&mut self) -> FrameRegisters{
        FrameRegisters::new(self.values.as_mut_ptr())
    }
}
\end{lstlisting}
The \TT{Vec<UntypedValue>} on line 2 is converted into a raw pointer on line 11. Once the \TT{Vec} is dropped, the raw pointer within the \TT{FrameRegisters} struct will point to freed memory. However, all the associated methods of the \TT{FrameRegisters} struct that dereference this pointer are marked \TT{unsafe}, which means that this vulnerability is not exploitable and can only be triggered with \TT{unsafe} code. Nevertheless, a small change to one of those methods could break this.

\subsection{User-implemented Reference Counting}
These are cases where the developer has bypassed the Rust lifetime ownership system in favour of implementing their own system of reference counting and memory management. Every time a pointer is duplicated or dropped, an internal reference count within the memory location is incremented or decremented. The memory is freed only if the count has reached zero. If any of these increments, decrements or checks are omitted, this could result in a use-after-free or memory leak. Here is an example from \TT{dbus}.

\lstset{xleftmargin=0.08\columnwidth, xrightmargin=0.05\columnwidth, escapeinside={<@}{@>}}
\begin{lstlisting}[caption=A code smell pattern in \TT{dbus}, captionpos=b, language=Rust, frame=none, style=colouredRust, basicstyle=\scriptsize \ttfamily, numbers=left]
pub struct Message {
  msg: *mut ffi::DBusMessage,
}
impl Message {
  pub fn new_method_return(m: &Message) -> Option<Message>{
    let ptr = unsafe { ffi::dbus_message_new_method_return(m.msg) };
    if ptr.is_null() { None } else { Some(Message { msg: ptr} ) }
  }
}
impl Drop for Message {
  fn drop(&mut self) {
    unsafe {
      ffi::dbus_message_unref(self.msg);
    }
  }
}
\end{lstlisting}
The pointer \TT{msg: *mut ffi::DBusMessage} is duplicated between two different \TT{Message} structures, but the duplication increments a reference counter through \TT{ffi::dbus\_message\_new\_method\_return}, and the \TT{Drop} implementation decrements this counter within \TT{ffi::dbus\_message\_unref}. This balance ensures that the pointer never becomes invalid while it is still accessible.

\subsection{Memory copy, not alias}
\sysname{} predicts that there could be a vulnerability if two pointers were aliased, however the function actually copies the underlying memory of the first pointer to create the target of the second pointer. Here is one such function from the \TT{napi} crate.

\lstset{xleftmargin=0.08\columnwidth, xrightmargin=0.05\columnwidth, escapeinside={<@}{@>}}
\begin{lstlisting}[caption=A code smell pattern in \TT{napi}, captionpos=b, language=Rust, frame=none, style=colouredRust, basicstyle=\scriptsize \ttfamily, numbers=left]
pub struct $name {
  data: *mut $rust_type,
  ...
}
impl $name {
  pub fn with_data_copied<D>(data: D) -> Self
  where
    D: AsRef<[$rust_type]>,{
    let mut data_copied = data.as_ref().to_vec();
    let ret = $name {
      data: data_copied.as_mut_ptr(),
      ...
    };
    mem::forget(data_copied);
    ret
  }
}
\end{lstlisting}
The type \TT{D} is a reference to a slice of some type \TT{\$rust\_type}. This underlying memory is then \textit{copied} by the function \TT{as\_ref()} on line 9, and a raw pointer to this copied memory is then transferred into the newly created structure. \sysname{} sees the \TT{as\_ref()} function as opaque, so it thinks that it is plausible that it is an alias, not a copy. However this does indicate a need for caution in the future development of this code. If the portion \TT{.as\_ref().to\_vec()} is removed from line 9, the memory would not be copied and this would be an exploitable vulnerability.

\subsection{Different field of structure}
\sysname{} predicts that there could be a vulnerability if two pointers were aliased, however the aliasing actually happens with a different field of a structure. All such examples from \sysname{} reports are fairly complicated to describe and involve nested structures, nevertheless we attempt to present one example from the \TT{rio} crate.
\lstset{xleftmargin=0.08\columnwidth, xrightmargin=0.05\columnwidth, escapeinside={<@}{@>}}
\begin{lstlisting}[caption=A code smell pattern in \TT{rio}, captionpos=b, language=Rust, frame=none, style=colouredRust, basicstyle=\scriptsize \ttfamily, numbers=left]
pub struct Uring {
  sq: Mutex<Sq>,
  ...
}
impl Uring {
  pub fn connect<'a, F>(
          &'a self,
          socket: &'a F,
          address: &std::net::SocketAddr,
          order: Ordering,
      ) -> Completion<'a, ()>
      where
          F: AsRawFd,
  ...
}
pub(crate) struct Sq {
  khead: *mut AtomicU32,
  ...
  array: &'static mut [AtomicU32],
}
pub struct Completion<'a, C: FromCqe> {
  ...
  uring: &'a Uring,
}
\end{lstlisting}
\sysname{} predicts that there could be a potential vulnerability because \TT{*(*(self).sq.0.khead)} outlives \TT{'a}, and it might be returned as \TT{*(*(ret.uring).sq.0.array)}, where it is supposed to live for the entire lifetime of the program (\TT{'static}). There is actually no connection between these two fields, which means that this is not a vulnerability. However the types of the two fields match, and it is not implausible that there could be an incorrect implementation of this function that connects these two fields, leading to an exploitable vulnerability.

\subsection{Two Lifetime Annotations}
Here is one example of this pattern from the \TT{inout} crate. It was the only one of its kind from the 85 reports that \sysname{} generated.

\lstset{xleftmargin=0.08\columnwidth, xrightmargin=0.05\columnwidth, escapeinside={<@}{@>}}
\begin{lstlisting}[caption=A code smell pattern in \TT{inout}, captionpos=b, language=Rust, frame=none, style=colouredRust, basicstyle=\scriptsize \ttfamily, numbers=left]
pub struct InOutBuf<'inp, 'out, T> {
  pub(crate) in_ptr: *const T,
  pub(crate) out_ptr: *mut T,
  pub(crate) len: usize,
  pub(crate) _pd: PhantomData<(&'inp T, &'out mut T)>,
}
impl<'inp, 'out, T> InOutBuf<'inp, 'out, T> {
  pub fn into_out(self) -> &'out mut [T]{
    unsafe { slice::from_raw_parts_mut(self.out_ptr, self.len) }
  }
}
\end{lstlisting}

The structure has two associated LAs, \TT{'inp} and \TT{'out}, corresponding to two raw pointers \TT{in\_ptr} and \TT{out\_ptr} within the structure. The raw pointer \TT{out\_ptr} is returned as a borrow with annotation \TT{'out}. But this is only valid if the underlying memory actually outlives \TT{'out}, and not \TT{'in}. This relies on the semantics of the functions that allocate memory for these pointers. In this case, that looks like this:

\lstset{xleftmargin=0.08\columnwidth, xrightmargin=0.05\columnwidth, escapeinside={<@}{@>}}
\begin{lstlisting}[caption=How memory is allocated for the raw pointers from the previous example, captionpos=b, language=Rust, frame=none, style=colouredRust, basicstyle=\scriptsize \ttfamily, numbers=left]
impl<'inp, 'out, T> From<(&'inp T, &'out mut T)>
                      for InOut<'inp, 'out, T> {
  #[inline(always)]
  fn from((in_val, out_val): (&'inp T, &'out mut T))
                                           -> Self {
    Self {
      in_ptr: in_val as *const T,
      out_ptr: out_val as *mut T,
      _pd: Default::default(),
    }
  }
}
\end{lstlisting}
Notice on line 4 that the first borrowed value outlives \TT{'inp} and the second value outlives \TT{'out}. This is \textit{enforced} when the structure is created, and is \textit{assumed} to be still true when the \TT{into\_out} function above is called. Developers must be careful to preserve this guarantee in all structure methods that manipulate these memory locations.

\pagebreak
\clearpage
 
\section{Extracting Lifetime Bounds}

\subsection{Notation}

\subsubsection{Grammar}
\Cref{fig:grammar} are the types that we work with, and lifetime annotations.

\begin{figure*}[t]
\begin{align*}
T ::&~\&L~\TT{mut}~T &\text{A mutable borrow to some type}\\
&|~\&L~T  &\text{An immutable borrow to some type}\\
&|~O &\text{A structure or a primitive type}\\
&|~\TT{*const}~|~\TT{*mut} &\text{(Im)mutable raw pointer to some type}\\
O::&~S &\text{A struct or primitive type not containing borrows}\\
&|~S\TT{<}L\TT{>} &\text{A struct containing borrows}\\
L ::&~\TT{'id}~|~\TT{'static} &\text{Lifetime names}\\
S ::&~\TT{id} &\text{Identifier names}
\end{align*}
\caption{Our grammar for types}
\label{fig:grammar}
\end{figure*}

\noindent
This surface syntax is almost identical to Rust. One important difference is that we assume that all borrow lifetimes are explicitly specified, whereas Rust allows some lifetimes to be inferred or \textit{elided}. Further, we've deliberately considered only structures with one lifetime annotation for simplicity, whereas Rust can have multiple.

\subsubsection{Judgements}

\Cref{fig:judgements} show the different kinds of judgements. Broadly, we want to process a value of a certain type and extract values of various subtypes from them. \textit{A key point is that we abstract away the details of the values themselves, and retain just their type information.} So a judgement like $T \rightarrow T ':~\epsilon$ would be interpreted as ``There is a value of type $T'$ that is reachable from the value of type $T$, and this value is owned by the parent value''. The actual details of which values these are are irrelevant to our analysis.

\begin{figure*}[t]
\begin{align*}
 &O\{T\}   \;\;\;  &\text{``}O \text{ contains a value of type } T \text{ as one of its fields''}\\
&T \rightarrow T ':~L  &\text{``} T \text{ contains another type T' that outlives lifetime } L \text{''}\\
&T \rightarrow T ':~\epsilon &\text{``} T \text{ contains another \textit{owned} type } T' \text{''}\\
&T \rightarrow L_1:~L_2 &\text{``From } T \text{, we can infer that } L_1 \text{ is longer than } L_2 \text{''}
\end{align*}
\caption{Judgements}
\label{fig:judgements}
\end{figure*}

\subsection{Rules for Decomposing a Type}

We want to take a type $T$ and extract all of its contained types with their associated lifetimes. We will do this using the rules in~\Cref{fig:rules}.

\bi
\item The simplest case - any owned type contains itself and owns itself (\TT{contains-self}).
\item If there is a borrow to an owned type, then it has to live at least as long as the lifetime of that borrow (\TT{borrow} and \TT{mut-borrow}).
\item If a structure has a field of type $T$, then anything that can be extracted from that $T$ can also be extracted from the structure (\TT{field} and \TT{field-eps}).
\item If an extracted type already has an associated lifetime, then any outer borrow lifetime does not attach to it (\TT{inner-lifetime}).
\item Any outer borrow lifetime must be shorter than an inner borrow lifetime (\TT{B-inner-longer}).
\item We can always conclude that a lifetime outlives itself\\ (\TT{B-reflexive}).
\item The static lifetime\footnote{\url{https://doc.rust-lang.org/rust-by-example/scope/lifetime/static_lifetime.html}} always outlives any other lifetime\\ (\TT{B-static}).
\item Lifetime bounds can also be extracted from inner types\\(\TT{B-extract-inner1} and \TT{B-extract-inner2}).
\ei

\begin{figure*}
\begin{mathpar}
\infer[\TT{contains-self}]{O \rightarrow O : \epsilon}{}
\and 
\infer[\TT{borrow}]{\&LT \rightarrow T_1:L}{T \rightarrow T_1 : \epsilon}
\and
\infer[\TT{mut-borrow}]{\&L~\TT{mut}~T \rightarrow T_1:L}{T \rightarrow T_1 : \epsilon}
\and
\infer[\TT{field}]{O \rightarrow T_1 : L}{O\{T\} & T \rightarrow T_1 : L}
\and
\infer[\TT{field-eps}]{S \rightarrow T_1 : \epsilon}{O\{T\} & T \rightarrow T_1 : \epsilon}
\and
\infer[\TT{inner-lifetime}]{\&LT \rightarrow T_1 : L_1}{T \rightarrow T_1 : L_1}
\and \\
\infer[\TT{raw-owned}]{S \rightarrow T: \epsilon }{S\{\TT{*const}~T\}}
\and
\infer[\TT{raw-mut-owned}]{S \rightarrow T: \epsilon }{S\{\TT{*mut}~T\}}
\and
\infer[\TT{raw-lifetime}]{S\TT{<}L\TT{>} \rightarrow T: L }{S\TT{<}L\TT{>}\{\TT{*const}~T\}}
\and
\infer[\TT{raw-mut-lifetime}]{S\TT{<}L\TT{>} \rightarrow T: L }{S\TT{<}L\TT{>}\{\TT{*mut}~T\}}
\and
\infer[\TT{B-inner-longer}]{\&LT \rightarrow L_1 : L}{T \rightarrow T_1 : L_1}
\and
\infer[\TT{B-reflexive}]{T \rightarrow L :~L}{}
\and
\infer[\TT{B-static}]{T \rightarrow \TT{'static} :~L}{}
\and
\infer[\TT{B-extract-inner1}]{T \rightarrow L_1 :~L_2}{T \rightarrow T_1 : L & T_1 \rightarrow L_1:~L_2}
\and
\infer[\TT{B-extract-inner2}]{T \rightarrow L_1 :~L_2}{T \rightarrow T_1 : \epsilon & T_1 \rightarrow L_1:~L_2}
\end{mathpar}
\caption{Rules for Extracting Lifetime Relationships. The first 6 rules deal with derivation of type-lifetime relationships for borrows and structures; the next 4 (starting with \TT{raw-}) are a generalization of the type-lifetime rules for borrows, to raw pointers; the next 5 (starting with \TT{B-}) deal with the calculation of implicit lifetime-lifetime relationships.}
\label{fig:rules}
\end{figure*}

\subsection{In the Presence of Raw Pointers}

In an ideal world, we want a structure's annotation to be representative of what it holds.

\bi
\item If a structure is annotated with lifetime \TT{<'a>}, then we would like everything accessible from the structure to outlive \TT{'a}.
\item If a structure has no lifetime annotation, then we would like everything accessible from the structure to be owned by the structure object.
\ei
\noindent
But raw pointers break that guarantee. A structure can have lifetime annotation \TT{<'a>}, but the thing pointed to by the raw pointer can live shorter than that. We want to \textit{enforce these guarantees}. We extend our rules as follows:
\bi
\item If a structure with no lifetime annotations contains a raw pointer, then the pointed-at type acts as though it is owned by the structure object (\TT{raw-lifetime} and \TT{raw-mut-lifetime}).
\item If a structure with a lifetime annotation \TT{<'a>} contains a raw pointer, then the data that it points to must outlive \TT{'a} (\TT{raw-owned} and \TT{raw-mut-owned}).
\ei

\section{Checking Violations}

\subsection{Arg-Return Use-after-free}
Suppose an argument and return type are of types $T_1$ and $T_2$ respectively. Then for all $L_1, L_2$ such that $T_1 \rightarrow T':L_1$ and $T_2 \rightarrow T':L_2$, if $T_1 \rightarrow L_1:L_2$ is not true, this is potentially Undefined Behavior (UB). In notation,

\begin{mathpar}
\infer[\TT{uaf1}]{\TT{UB}}{T_1 \rightarrow T':L_1 & T_2 \rightarrow T':L_2 & T_1 \nrightarrow L_1:L_2} \and
\infer[\TT{uaf2}]{\TT{UB}}{T_1 \rightarrow T':L_1 & T_2 \rightarrow T':\epsilon}
\end{mathpar}

Here we abuse notation by using judgement $T_1 \nrightarrow L_1:L_2$ to mean that we cannot extract $L_1:L_2$ from $T_1$. We do \textit{not} check if $T_2 \nrightarrow L_1:L2$, because we can only \textit{infer} constraints from the arguments we are given. We \textit{impose} constraints on the return types.

We report violations to our alias analysis module that checks if there is \textit{actually a transfer} of a value of type $T'$ between $T_1$ and $T_2$.

\subsection{Arg-Arg Use-after-free}
Suppose two arguments are of types $T_1$ and $T_2$. Then, the rules for checking Arg-Arg Use-after-free violations are:

\begin{mathpar}
\infer[\TT{uaf3}]{\TT{UB}}{T_1 \rightarrow T':L_1 & T_2 \rightarrow T':L_2 & T_1 \nrightarrow L_1:\TT{'static}}
\end{mathpar}

Any non-static lifetime being assigned to another lifetime, between matching types, gets flagged as a violation, to be taken care of by the alias analyzer.

\subsection{Arg-Return Non-Exclusive Mutability}
Suppose an argument and return type are of types $T_1$ and $T_2$ respectively. Then, the rules for checking Non-Exclusive Mutability (NEM) violations are:

\begin{mathpar}
\infer[\TT{nem1}]{\TT{UB}}{T_1 \rightarrow T':L_1 & T_2 \rightarrow \&\TT{mut}~T':L_2 & T_1 \nrightarrow L_1:L_2} \and
\infer[\TT{nem2}]{\TT{UB}}{T_1 \rightarrow T':L_1 & T_2 \rightarrow \TT{*mut}~T':L_2 & T_1 \nrightarrow L_1:L_2} \and
\infer[\TT{nem3}]{\TT{UB}}{T_1 \rightarrow \TT{*mut}~T':L_1 & T_2 \rightarrow T':L_2 & T_1 \nrightarrow L_1:L_2} \and
\infer[\TT{nem4}]{\TT{UB}}{T_1 \rightarrow \&\TT{mut}~T':L_1 & T_2 \rightarrow T':L_2 & T_1 \nrightarrow L_1:L_2}
\end{mathpar}

Again, we report violations to our alias analysis module that checks if there is \textit{actually a transfer} of a value of type $T'$ between $T_1$ and $T_2$.

\pagebreak
\clearpage
\section{Alias Analysis}

This section describes our alias analysis in more detail. For a variable $a$, we denote its points-to set by $S_a$. For dereferencing, we define $*S_a = \bigcup_{x \in S_a} S_x$. We process each statement in order, and for each type of assignment statement, like \TT{a = b}, \TT{a = *b}, etc, we update the points-to sets accordingly. If there is a function call, then we consider all possible kinds of assignments and dereferences between the function arguments, like \TT{arg1 = *arg2}, \TT{ **arg1 = arg2}, etc. The pseudocode is shown in~\Cref{alg:alias_analysis}. Note that this is a simplified version that does not include field sensitivity. For the full algorithm, please refer to our code on GitHub.

\begin{algorithm}[h]
\begin{lstlisting}[language=algo, frame=none, style=normal]
input: body
input: source, target

source_val := alloc()
$S_{\text{source}}$ := { source_val }

for basic_block in reverse_postorder(body):
    for statement in basic_block.statements:
        match statement
            [a = b] =>
                $S_a := S_a \cup S_b$
                
            [*a = b] =>
                if $S_a$ is empty:
                    $S_a := \{~\texttt{alloc()}~\}$
                for $x \in *S_a$:
                    $S_x := S_x \cup S_b$
                    
            [a = *b] =>
                if $S_b$ is empty:
                    $S_b := \{~\texttt{alloc()}~\}$
                for $x \in *S_b$:
                    $S_a := S_a \cup S_x$

            [a = &b] =>
                $S_a := S_a \cup \{~b~\}$

            [ret = f(args)] =>    
                for a, b $\in$ choose(args, 2):
                    for $x \in (S_a \cup *S_a \cup **S_a ...)$:
                        for $y \in (S_b \cup *S_b \cup **S_b ...)$:
                            $S_x := S_x \cup S_y$
                            $S_y := S_x \cup S_y$
                for a in args:
                    $S_\text{ret} := S_\text{ret} \cup S_a \cup *S_a \cup **S_a ...$

if $\texttt{source\_val} \in S_\texttt{target}$:
    return True // May Alias
else:
    return False // Not Aliased
\end{lstlisting}
\caption{Alias Analysis}
\label{alg:alias_analysis}
\end{algorithm}

\clearpage
\section{Shallow Filter}

In this section, we describe some details about our shallow filter. This is based on some primitive inter-procedural analysis, as well as knowledge of common patterns of false positives.

\subsection{Filter based on \TT{Drop} Implementations}

Consider the hypothetical function \TT{to\_bar} shown in Listing \ref{lst:dropimpl}.

\begin{lstlisting}[language=Rust, frame=none, style=colouredRust, basicstyle=\footnotesize \ttfamily, numbers=left, captionpos=b, label=lst:dropimpl, caption=Example to illustrate filtering based on absence of \TT{Drop} implementations.]
pub struct Foo {
    data: *const i32
}
pub struct Bar {
    data: *const i32
}
impl Foo {
    pub fn new(val: i32) -> Foo {
        Foo{data: Box::leak(Box::from(val))}
    }
    pub fn to_bar(&self) -> Bar {
        Bar{data: self.data}
    }
    pub fn get_val(&self) -> &i32 {
        unsafe{&*self.data}
    }
}
\end{lstlisting}

\noindent
Consider the function \TT{to\_bar}. Our analysis would derive that \TT{self.data} must outlive the anonymous lifetime on \TT{\&self}, and when it is transferred over to the returned \TT{Bar} structure, it is valid for as long as the receiver (the function caller) holds on to it. This would be marked as a potential error, because our analysis cannot assume that \TT{self.data} will be valid for the entire lifetime of the program. However, in this program, that memory is leaked and doesn't ever get freed. A memory leak is \textbf{not} considered an exploitable vulnerability or undefined behavior in \rust.

\noindent
Hence we implement a simple inter-procedural check. If the value being re-assigned is behind a raw pointer in \textbf{both} the source as well as the target structures, then we only mark this as a violation if \textbf{either} of those structures has a custom \TT{Drop} implementation. A custom \TT{Drop} implementation would usually involve cleaning up memory, and this means that when that structures gets dropped, the \TT{data} value might be freed.

\subsection{Filter Common Patterns of False Positives}

There are certain common traits in \rust that rely on the semantic correctness of their implementation to ensure their safety. One such example is the \TT{IteratorMut} trait, specifically its \TT{next} and \TT{next\_back} methods \cite{splitting_borrows}. Listing \ref{lst:iternext} shows an example from the \TT{lru} crate.

\begin{figure}
\begin{lstlisting}[language=Rust, frame=none, style=colouredRust, basicstyle=\scriptsize \ttfamily, numbers=left, captionpos=b, label=lst:iternext, caption=An example from \TT{lru} showing a common pattern of false positive.]
impl<'a, K, V> Iterator for IterMut<'a, K, V> {
  type Item = (&'a K, &'a mut V);

  fn next(&mut self) -> Option<(&'a K, &'a mut V)> {
    // Returns a reference to value at index
    // Moves pointer ahead by 1
  }
}
\end{lstlisting}
\end{figure}

This is a mutable iterator which allows mutable access to the elements of a cache. The lifetime of the returned value \TT{(\&’a mut V)} and the lifetime of \TT{\&mut self} are unrelated, which means that it is possible to call \TT{next()} multiple times to get multiple mutable references to elements of the cache. However, the semantics of the function ensure that each of these returned references always points to a different cache element. Once a value has been returned, the iterator moves to the next value and there is no way to “rewind” the iterator to point back to the previous value. If we had another function \TT{rewind()} that moves the pointer back by one entry, then we could use \TT{rewind()} and \TT{next()} together to create two aliased mutable references as follows:

\begin{lstlisting}[language=Rust, frame=none, style=colouredRust, basicstyle=\footnotesize \ttfamily, numbers=left, captionpos=b]
let (k1, v1) = iter.next();
iter.rewind(); // Moves pointer back by 1
let (k2, v2) = iter.next();
// v1 and v2 would be aliased mutable references
\end{lstlisting}

However the \TT{rewind()} function doesn’t exist, and so the \TT{next()} function can always be called safely. There is no way to predict this without reasoning about the semantic properties of the code.

The \TT{Iterator} trait is implemented for different types in many crates, and in each of these, the \TT{next} function bears a similar signature and implementation semantics. \sysname{} flags this as a potential violation in several of these cases. To remedy this problem, our shallow filter simply ignores implementations of \TT{Iterator::next} altogether.

There are three specific functions from trait implementations that we filter out - \TT{Iterator::next}, \TT{Iterator::next\_back}, \TT{Clone::clone}. It is true that there can be potential flaws in these implementations, which is why the shallow filter is the last stage of our analysis, and a user has the option of turning it off.

\end{document}